\documentclass[acmtog]{acmart}
\usepackage{setspace,afterpage,fix-cm,multirow}
\usepackage{amssymb}
\usepackage{enumitem}
\setitemize{noitemsep,topsep=0pt,parsep=0pt,partopsep=0pt,leftmargin=*}
\usepackage{hyperref}
\usepackage{pdfpages}
  
\usepackage[noend]{algpseudocode}
\usepackage{booktabs}
\usepackage{mathtools}
\usepackage{multirow}
\usepackage{tabstackengine}
\usepackage{subcaption}
\usepackage{siunitx}
\usepackage{bm}

\mathtoolsset{centercolon}

\newcommand{\mydraft}{false}

\newcommand{\FF}{\mathbf{F}}

\newcommand{\PP}{\mathbf{P}}

\newcommand{\HH}{\mathbf{H}}

\newcommand{\DD}{\mathbf{D}}

\newcommand{\ff}{\mathbf{f}}

\usepackage{pifont}

\newcommand{\bb}{\mathbf{b}}

\renewcommand{\gg}{\mathbf{g}}

\newcommand{\xx}{\mathbf{x}}

\newcommand{\vv}{\mathbf{v}}

\usepackage[ruled]{algorithm2e} 

\SetAlFnt{\small}
\SetAlCapFnt{\small}
\SetAlCapNameFnt{\small}
\SetAlCapHSkip{0pt}
\IncMargin{-\parindent}
\acmJournal{TOG}
\setcitestyle{square}

\makeatletter
\renewcommand{\l@section}{\@dottedtocline{1}{1.5em}{2.6em}}
\renewcommand{\l@subsection}{\@dottedtocline{2}{4.0em}{3.6em}}
\renewcommand{\l@subsubsection}{\@dottedtocline{3}{7.4em}{4.5em}}
\makeatother

\begin{document}

\title{Hierarchical Optimization Time Integration for CFL-rate MPM Stepping}
\author{Xinlei Wang}
\authornote{equal contributions}
\affiliation{%
  \institution{Zhejiang University \& University of Pennsylvania}
  }
\email{wxlwxl1993@zju.edu.cn}

\author{Minchen Li}
\authornotemark[1]
\affiliation{%
  \institution{University of Pennsylvania \& Adobe Research}
  }
\email{minchernl@gmail.com}

\author{Yu Fang}
\affiliation{%
  \institution{University of Pennsylvania}
  }
\email{squarefk@gmail.com}

\author{Xinxin Zhang}
\affiliation{%
  \institution{Tencent}
  }
\email{zhangshinshin@gmail.com}

\author{Ming Gao}
\affiliation{%
  \institution{Tencent \& University of Pennsylvania}
  }
\email{ming.gao07@gmail.com}

\author{Min Tang}
\affiliation{%
  \institution{Zhejiang University}
  }
\email{tang_m@zju.edu.cn}

\author{Danny M. Kaufman}
\affiliation{%
  \institution{Adobe Research}
  }
\email{kaufman@adobe.com}

\author{Chenfanfu Jiang}
\affiliation{%
  \institution{University of Pennsylvania}
  }
\email{cffjiang@seas.upenn.edu}

\begin{abstract}

We propose Hierarchical Optimization Time Integration (HOT) for efficient implicit timestepping of the Material Point method (MPM) irrespective of simulated materials and conditions. HOT is a MPM-specialized hierarchical optimization algorithm that 
solves nonlinear time step problems for large-scale MPM systems near the CFL-limit.
HOT provides convergent simulations ``out-of-the-box'' across widely varying materials and computational resolutions without parameter tuning. 
  As an implicit MPM timestepper accelerated by a custom-designed Galerkin multigrid wrapped in a quasi-Newton solver, HOT is both highly parallelizable and robustly convergent. As we show in our analysis, HOT maintains consistent and efficient performance even as we grow stiffness, increase deformation, and vary materials over a wide range of finite strain, elastodynamic and plastic examples. Through careful benchmark ablation studies, we compare the effectiveness of HOT against seemingly plausible alternative combinations of MPM with standard multigrid and other Newton-Krylov models. We show how these alternative designs result in severe issues and poor performance. In contrast, HOT outperforms existing state-of-the-art, heavily optimized implicit MPM codes with an up to 10$\times$ performance speedup across a wide range of challenging benchmark test simulations.

\end{abstract}

%
%
%
%


\maketitle

\section{Introduction} \label{sec:intro}

\begin{figure}
\centering
   \includegraphics[draft=false,width=\linewidth]{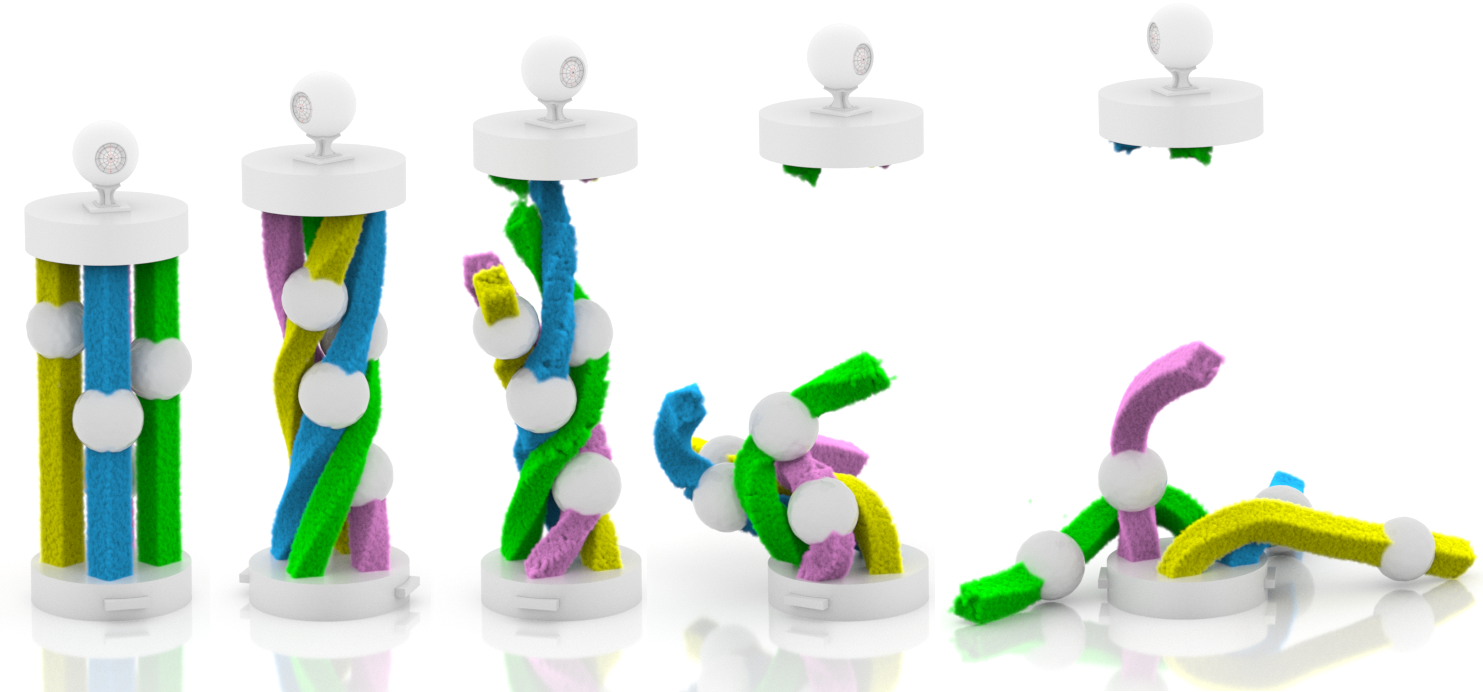}
   \caption{HOT is naturally suited for simulating dynamic contact of heterogeneous solid materials with substantial stiffness discrepancy. In this bar twisting example, compared across all available state-of-the-art, heavily optimized implicit MPM codes, HOT achieves more than 4$\times$ speedup overall and up to 10$\times$ per-frame. HOT obtains rapid convergence without need for per-example hand-tuning of either outer nonlinear solver nor inner linear solver parameters. \label{fig:bars}}
   \label{fig:twist}
\end{figure}


 The Material Point method (MPM) is a versatile and highly effective approach for simulating widely varying material behaviors ranging from stiff elastodynamics to viscous flows (e.g. Figures\ \ref{fig:wheel} and  \ref{fig:sauce}) in a common framework.
 %
As such MPM offers the promise of a single unified, consistent and predictive solver for simulating continuum dynamics across diverse and potentially heterogenous materials. However, to reach this promise, significant hurdles remain.
Most significantly, obtaining accurate, consistent and robust solutions within a practical time budget is severely challenged by small timestep restrictions. This is most evidenced as we vary material properties, amounts of deformation and/or simulate heterogenous systems (see Table \ref{table:real-materials}). 

While MPM's Eulerian grid resolution limits time step sizes to the CFL limit\footnote{A particle cannot travel more than one grid cell per time step while, in practice, a CFL number of $0.6$ is often used \cite{gast:2015:tvcg}.}\ \cite{fang2018temporally}, the explicit time integration methods commonly employed for MPM often require much smaller time steps. In particular, the stable timestep sizes of explicit MPM time integration methods remain several orders of magnitude below the CFL limit when simulating stiff materials like metal (see Table~\ref{table:real-materials}) and snow\ \cite{stomakhin:2013:snow,fang2018temporally}.
A natural solution then is to apply implicit numerical time integration methods, i.e. implicit Euler, which can enable larger stable time step sizes for MPM\ \cite{gast:2015:tvcg,Fang:2019:ViscousMPM}. However, doing so requires solving challenging and potentially expensive nonlinear systems at every timestep.

\subsection{Challenges to implicit MPM timestepping}
While implicit MPM timestepping methods in engineering provide larger step sizes\ \cite{guilkey2001implicit,cummins2002implicit,guilkey:2003:implicit-mpm,nair2012implicit}, they do not target CFL-rate time step sizes as is typically desired in graphics applications. Thus, in engineering, the standard Newton method is often directly applied without globalizations to solve the nonlinear timestepping problem\ \cite{nair2012implicit, charlton2017igimp}; with larger time step sizes near the CFL limit, the nonlinearity grows and the 1st-order Taylor expansion becomes less accurate, which can then make Newton method unstable and even explode.
More recently state-of-the-art implicit MPM methods in graphics have been introduced that enable time steps closer to the CFL limit. Gast and colleagues\ \shortcite{gast:2015:tvcg} introduced a globalized Newton-Krylov method for MPM while Fang et al.\ \shortcite{Fang:2019:ViscousMPM} extended ADMM to solve implicit MPM timesteps.
However, their convergence and performance are limited for simulations involving heterogeneous and/or stiff materials, leading to slow computations and inconsistent, unpredictable and even unstable results.
%
While ADMM\ \cite{Fang:2019:ViscousMPM} for MPM is attractively efficient, the underlying ADMM algorithm has no guarantee of convergence for non-convex and nonlinear continua problems. In practice it can at best achieve linear convergence. As we show in Section \ref{sec:results}, when able to converge the ADMM solver is thus exceedingly slow to reach reasonable solutions.
 
On the other hand inexact Newton-Krylov methods exemplified by Gast et al.\ \shortcite{gast:2015:tvcg} are seemingly ideal for solving implicit MPM problems where the sparsity structure of the Hessian can change at every timestep. Key to the efficiency and stability of these methods are the \emph{inexact} iterative linear solve of each inner Newton iterate. In turn this requires setting a tolerance to terminate each such inner loop. 
However, no single tolerance setting works across examples. Instead, suitable tolerances can and will vary over many orders of magnitude per example and so must be experimentally determined as we change set-ups over many expensive, successive simulation trials. Otherwise, as we demonstrate in Section \ref{sec:results} and our supplemental document, a tolerance suitable for one simulated scene will generate extremely slow solves, non-physical artifacts, instabilities and even explosions in other simulations. 

Next, we observe that 
Newton-Krylov methods employing Jacobi or Gauss-Seidel preconditioned CG solvers significantly lose efficiency from deteriorating convergence as material  stiffnesses increase (Table\ \ref{tb:sgs_jacobi}). 
In such cases multigrid preconditioners\ \cite{wang:2018:multigridcloth, Tamstorf:2015:multigridcloth, zhu2010efficient} are often effective solutions as the underlying hierarchy allows aggregation of multiple approximations of the system matrix inverse across a range of resolutions. This  accelerates information propagation across the simulation domain and thereby improves convergence.

We focus on h-multigrid that coarsens the degrees of freedom for coarser levels to reduce the computational cost. 
H-multigrid has been investigated for MPM by Cummings and Brackbill\ \shortcite{cummins2002implicit} via merging particles level by level. However, they conclude that it performs similarly to Jacobi preconditioners, which indicates that building a multigrid hierarchy for MPM is challenging.
This may be because merging particles lacks error bounds and can potentially make DOF coarsening inconsistent.
However, as we discuss in Section\ \ref{subsec:baseline}, although building each coarser level system using the original particles without merging improves convergence of inner linear solves, the computational overhead of this seemingly reasonable hierarchy still does not reduce the overall cost of MPM simulations. 
This is because (1) construction and evaluation of system matrices in each coarser level can be as expensive as the fine level computation, and (2) at the domain boundaries, the coarsening of DOFs may not be consistently defined between matrices and right-hand side vectors of the coarse level systems. 

\subsection{Hierarchical Optimization Time Integration}

We propose the HOT algorithm to address these existing limitations and so provide ``out of the box'' efficient MPM simulation.
 To enable consistent, automatic termination of both outer Newton iterations and inner inexact linear solves across simulation types we extend the characteristic norm\ \cite{Li:2019:DOT,bcqn18} to inhomogenous MPM materials. As we show in Section\ \ref{sec:CN}) and Table 1 in our supplemental document, this produces consistent, automatic, high-quality results for inexact Newton-Krylov simulations that match the quality and timing of the best hand-tuned results of Gast et al.\ \shortcite{gast:2015:tvcg}. 
 
Next, to obtain both improved convergence \emph{and} performance for MPM systems with multigrid we develop a new, MPM-customized hierarchy. We begin by embedding progressively finer level grid nodes into coarser level nodes via the MPM kernel.  We then construct coarse level matrices directly from their immediate finer level matrix entries. This avoids computation and storage of each coarse level's geometric information, automatically handles boundary conditions and enables sparsity by our choice of MPM embedding kernel. This resulting multigrid hierarchy then retains improved convergence while also significantly improving performance; see Figure\ \ref{fig:baseline}.
 
While offering a significant gain, our MPM-customized multigrid still requires explicit matrix construction. In many elastodynamic simulation codes, such matrix construction costs are alleviated by applying just a fixed number of Newton iterations irrespective of convergence. However this strategy is neither suitable for artistic control nor engineering as it sacrifices consistency and accuracy for efficiency; e.g., it can produce artificially softened materials, numerically damped dynamics and inaccurate predictions. 
Following recent developments in mesh-based elasticity methods\ \cite{Li:2019:DOT} we instead alleviate matrix construction costs by constructing our hierarchy just once per timestep  but then apply it as an efficient, second-order initializer (with one V-cycle per iteration) inside a quasi-Newton solver. 

\subsection{Contributions}

HOT's inner multigrid provides efficient second-order information, while its outer quasi-Newton low-rank updates provide efficient curvature updates. This enables HOT to maintain consistent, robust output with a significant speedup in performance -- even as we grow stiffness, increase deformation and widely vary materials across the simulation domain. 
The combined application of node-embedding multigrid, automatic termination, and customized integration of multigrid V-cycle into the quasi-Newton loop jointly enable HOT's significant and consistent performance gains. In summary, our contributions are


\begin{itemize} 
\item We derive a novel MPM-specific multigrid model exploiting the regularity of the background grid and construct a Galerkin coarsening operator consistent with re-discretization via particle quadrature. To our knowledge, this is the \emph{first time} Galerkin h-multigrid is applied for the MPM discretization of nonlinear elasticity with significant performance gain.
\item We develop a new, node-wise Characteristic Norm\ \cite{Li:2019:DOT,bcqn18} (CN) measure for MPM. Node-wise CN enables unified tolerancing across varying simulation resolutions, material parameters and heterogenous systems for both termination of inner solves in inexact Newton and convergence determination across methods. CN likewise ensures a fair comparison across all solvers in our experiments.
\item We construct HOT -- an \emph{out-of-the-box} implicit MPM time integrator by employing our multigrid as an efficient inner initializer inside a performant quasi-Newton MPM time step solve. A carefully designed set of algorithmic choices customized for MPM then achieve both efficiency and accuracy that we demonstrate on a diverse range of numerically challenging simulations.
\item We perform and analyze extensive benchmark studies on challenging, industrial scale simulations to determine these best data structure and algorithmic choices for MPM numerical time integration. Across these simulation examples, we compare HOT against a wide range of alternative, seemingly reasonable algorithmic choices to demonstrate their pitfalls and the carefully designed advantages of HOT.
\end{itemize}

Across a wide range of challenging elastodynamic and plastic test simulations we show (see Section \ref{sec:results}) that HOT \emph{without the need of any parameter tuning} outperforms existing state-of-the-art, heavily optimized implicit MPM codes. All alternative methods either exhibit significantly slower performance or else suffer from large variations across simulated examples. Our study then suggests HOT as a robust, unified  MPM time integrator with fast convergence and outstanding efficiency of up to $10\times$ speedup to best alternatives across a wide range of possible simulation input.

\begin{table}[b]
\centering
\fontsize{.72em}{.55em}\selectfont
\caption{Parameters for solid materials studied in this paper.
  \label{table:real-materials}
}
\begin{minipage}{\columnwidth}
\centering
\begin{tabular}{@{}c@{\,\,\,\,}c@{\,\,\,\,}c@{\,\,\,\,}c@{\,\,\,}c@{\,\,\,}}
\toprule
                              & Density ($kg/m^3$)   &  Young's modulus (Pa) & Poisson's ratio & Yield stress (Pa)   \\ \midrule
Tissue   &  $300-1000$      & $1\times 10^2-1\times 10^6$      & $0.4-0.5$ & - \\ 
Rubber   &  $1000-2500$      & $1\times 10^6-5\times 10^7$      & $0.3-0.5$ & - \\ 
Bone   &  $800-2000$      & $7\times 10^7-3\times 10^{10}$      & $0.1-0.4$ & - \\ 
PVC   &  $1000-2000$      & $2\times 10^9-4\times 10^9$      & $0.3-0.4$ & $1\times 10^{7}-4\times 10^{7}$ \\ 
Metal   &  $500-20000$      & $1\times 10^{10}-4\times 10^{11}$      & $0.2-0.4$ & $2\times 10^{8}-2\times 10^{9}$ \\ 
Ceramic   &  $2000-6000$      & $1\times 10^{11}-4\times 10^{11}$      & $0.2-0.4$ & - \\ 
\bottomrule
\end{tabular}
\end{minipage}
\end{table}

\begin{figure}[t]
    \centering
    \includegraphics[draft=\mydraft,width=\linewidth]{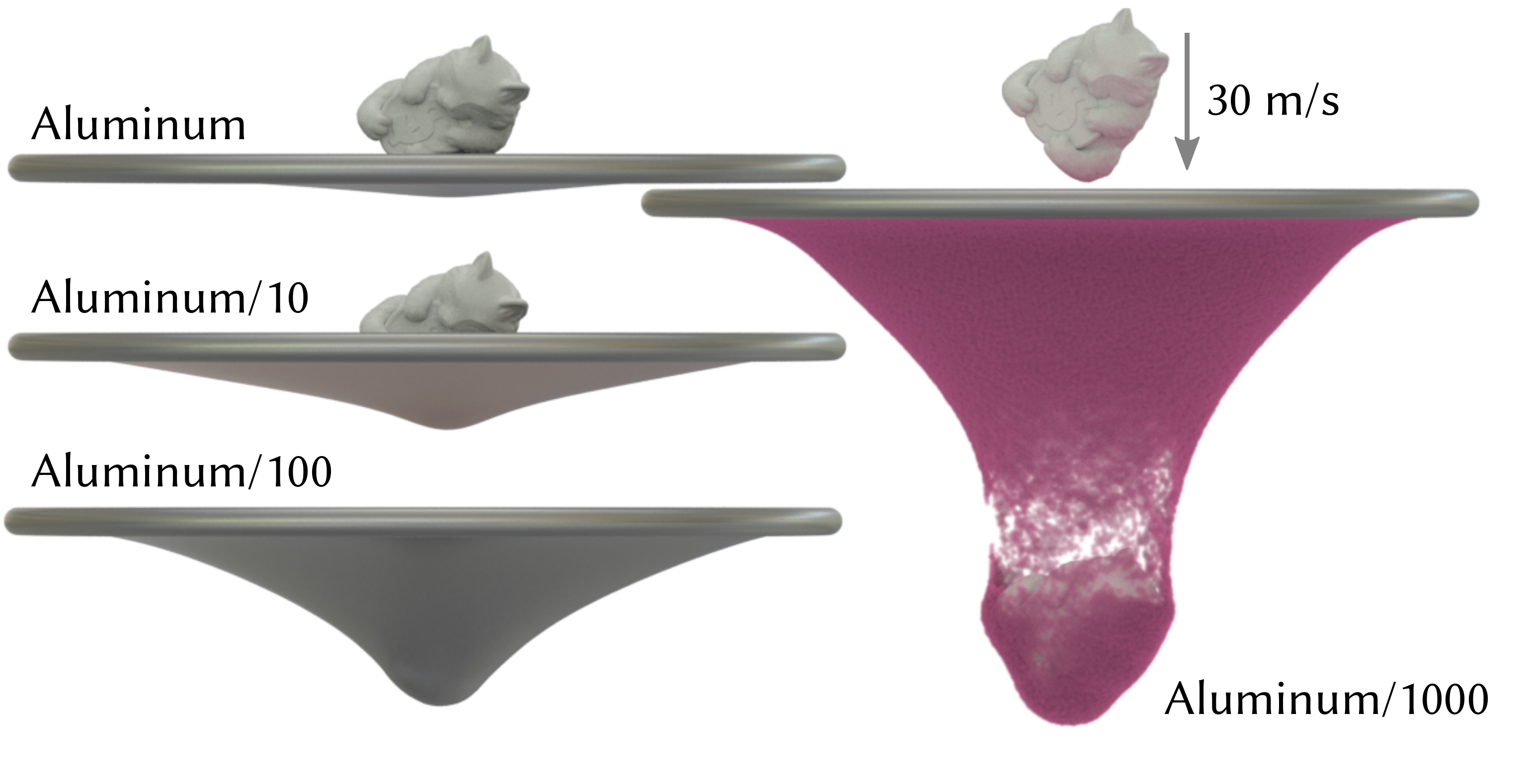}
    \caption{\textbf{Stiffness comparisons.} A stiff lucky cat is smashed onto sheets with different Young's moduli starting from aluminum (6.9Gpa, with yield stress 240Mpa) and then being scaled down by $10$, $100$ and $1000$. Different stiffness gives drastically different behavior for elatoplastic materials.}
    \label{fig:stiffness}
\end{figure}

\section{Related Work} \label{sec:related}
\subsection{Material Point Method} 
MPM was introduced by Sulsky et al. \shortcite{sulsky:1994:history-materials} as a generalization of FLIP \cite{brackbill1988flip,zhu:2005:sand-fluid} to solid mechanics. 
MPM's convergence was demonstrated computationally and explained theoretically by Steffen et al.\ \shortcite{steffen:2008:analysis} with a smooth, e.g. quadratic B-spline, basis for grid solutions.
This was further verified by Wallstedt\ \shortcite{wallstedt:2009:order} with manufactured solutions.

In graphics, MPM has been applied to model a diverse array of materials and behaviors. These include the modeling of snow \cite{stomakhin:2013:snow}, sand \cite{gao2018gpu,yue2018hybrid,daviet:2016:smp,klar:2016:des}, foam \cite{yue:2015:mpm-plasticity,ram:2015:foam,Fang:2019:ViscousMPM}, cloth \cite{jiang:2017:cloth,guo2018material}, rods\ \cite{han2019hybrid, fei2019mmc}, mixtures \cite{MS:2019}, fracture \cite{Wolper:2019:CDMPM,Wretborn:2017:ACP:3170009.3170083,wang2019simulation}, multiphase flow \cite{stomakhin:2014:augmented-mpm,andre:2017:wetsand,gao2018animating}, and even baking\ \cite{ding2019thermomechanical}. The coupling between softer MPM materials and rigid bodies has also been explored in both explicit \cite{hu2018} and implicit \cite{ding2019penalty} settings. While rigid body dynamics provides an efficient approximation of extremely stiff materials for many applications, it is not suitable for capturing elastoplastic yielding nor for computing accurate mechanical responses.

Implicit time integration, e.g., via implicit Euler, is often the preferred choice for timestepping stiff materials and large deformations due to explicit integration's often unacceptable sound-speed CFL restriction\ \cite{fang2018temporally}.
Early implicit MPM\ \cite{guilkey2001implicit, guilkey:2003:implicit-mpm} solutions applied Newmark time integration, demonstrating improved stability and solution accuracy over explicit MPM when compared to validated finite element solutions.
More recently, Nair and Roy\ \shortcite{nair2012implicit} and Charlton et al.\ \shortcite{charlton2017igimp} further investigated implicit generalized interpolation MPM for hyperelasticity and elastoplasticity respectively.
On the other hand, research in graphics has explored force linearization\ \cite{stomakhin:2013:snow} and optimization-stabilized Newton-Raphson solutions for both implicit Euler \cite{gast:2015:tvcg} and implicit midpoint\ \cite{jiang2017angular} to achieve larger time step sizes.
\begin{figure}[t]
    \centering
    \includegraphics[draft=\mydraft,width=\linewidth]{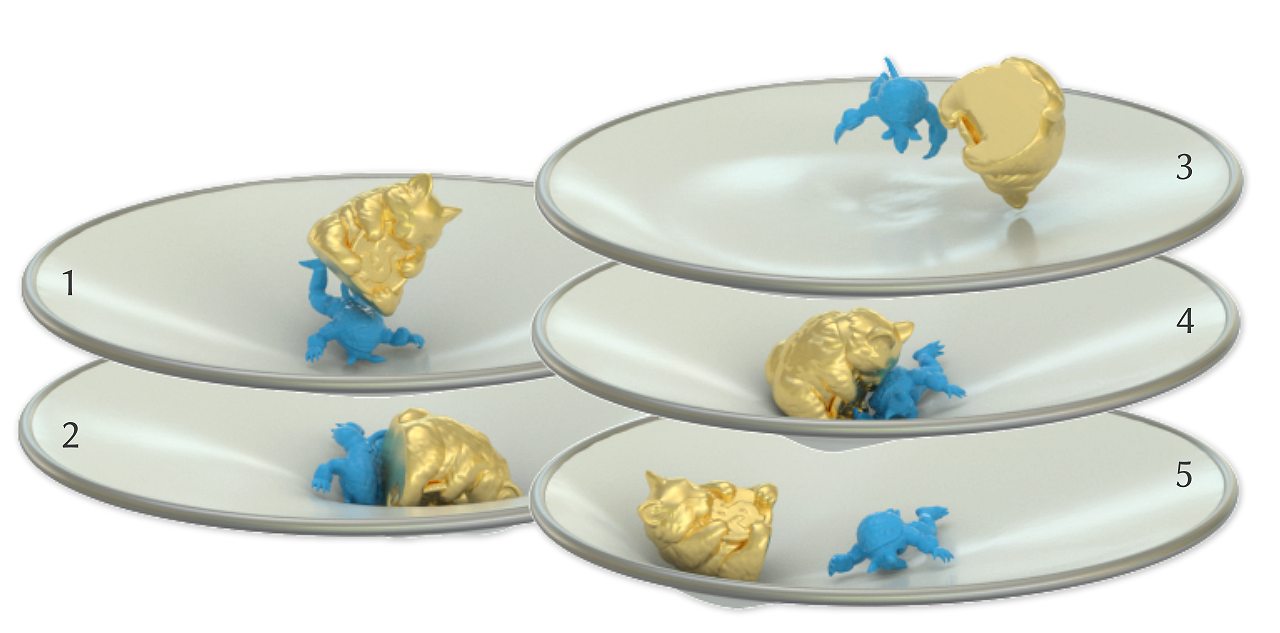}
    \caption{\textbf{ArmaCat.} A soft armadillo and a stiff lucky cat are both dropped onto an elastic trampoline, producing interesting interactions between them.}
    \label{fig:cat}
\end{figure}
\subsection{Optimization and non-linear integrators}
Numerical integration of differential systems is often reformulated variationally in order to be solved as a minimization problem. This allows methods to often achieve improved accuracy, robustness and performance by taking advantage of available numerical optimization techniques. 
In computer graphics, simulation methods are increasingly applying this strategy to simulate both fluid \cite{batty:2007:solid-fluid,Weiler2016-ud} and solid \cite{gast:2015:tvcg,wang2016descent,bouaziz2014projective,Overby2017-ns,Dinev2018-pp,Dinev2018-zn} dynamics.
For optimizations originating from nonlinear problems, Newton-type methods are generally the standard mechanism, delivering quadratic convergence near solutions. 
However, when the initial guess is far from a solution, Newton's method may fail to provide a reasonable search direction as the Hessian can be indefinite\ \cite{Li:2019:DOT,liu2017quasi,Smith2018-hv}.
Teran et al. \shortcite{teran2005robust} propose a positive definite fix to project the Hessian to a symmetric positive definite form to guarantee a descent direction can be found; we compare with this method and further augment it with a backtracking line search to ensure energy decrease.
We refer to this method as projected Newton (PN). 
Since PN requires a potential energy to be defined for the timestepping, we lag the plasticity update, and only perform it once per time step so that our system stays integrable. This is unlike Charlton et al.\ \shortcite{charlton2017igimp}, Kl\'ar et al.\ \shortcite{klar:2016:des}, and Fang et al.\ \shortcite{Fang:2019:ViscousMPM}, who handle plasticity fully implicitly.

In each PN iteration, a linear system is solved.
For MPM simulations which generally involve a large number of nodes and can have changing sparsity, Krylov iterative linear solvers such as conjugate gradient (CG) are generally preferred over direct factorization. 
To improve CG convergence, different preconditioning options exist. We apply the most efficient and straightforward Jacobi (diagonal) preconditioner as our baseline PN method which we refer as PN-PCG.
To further minimize memory consumption and access cost, existing implicit MPM methods in graphics apply matrix-free (MF) PN-PCG 
without explicitly constructing system matrices.
However, when many CG iterations are required, e.g., for large time step sizes and/or stiff materials, matrix-free is no longer necessarily a better option than matrix construction (Section\ \ref{sec:results}). This is because the cost of recomputing the intermediate variables becomes more dominant, while such a cost could be significantly reduced if the matrix is explicitly constructed only once at the beginning of the time step.
Convergence can then be further improved with multigrid preconditioning. However, doing so for MPM presents new challenges.

\subsection{Multigrid methods} \label{sec:related_multigrid}
Multigrid methods \cite{MGBook} are widely employed to accelerate 
both solid\ \cite{McAdams2011-kv,zhu2010efficient,wang:2018:multigridcloth,Tamstorf:2015:multigridcloth,xian2019multigrid,tielen2019efficient} and fluid\  \cite{fidkowski2005p,mcadams2010parallel,Setaluri:2014:spgrid,Zhang2015-lv,Zhang2016-hv,Zhang2014-io,gao2018animating,Aanjaneya2017-mi} dynamics simulations. 
Here multi-level structures allow information of computational cells to better propagate, making multigrid methods highly efficient for systems with long-range energy responses and/or high stiffnesses. 
Unlike p-multigrid\ \cite{fidkowski2005p, tielen2019efficient} methods which apply higher-order shape functions with same DOFs to improve convergence,
h-multigrid methods construct hierarchies of coarser DOF models with potentially lower computational cost.

H-multigrid is generally categorized as geometric or algebraic.
Unlike algebraic multigrid, 
geometric multigrid constructs coarse level system matrices from coarsened grids or meshes\ \cite{stuben2001review}.
However, the mismatch at the irregular boundaries due to geometric coarsening can require special treatment to ensure convergence improvement; e.g. extra smoothing at boundaries as in McAdam et al.\ \shortcite{mcadams2010parallel}. 
Alternately, Chentanez and M\"{u}ller \shortcite{nvidiaMGFlow} demonstrate that with a volume weighted discretization robust results can be obtained without additional smoothing at boundaries. 
On the other hand, Ando et al.\ \shortcite{ando2015pressure} derive a multi-resolution pressure solver from a variational framework which handles boundaries using signed-distance functions.
Cummings and Brackbill\ \shortcite{cummins2002implicit} proposed a geometric multigrid-preconditioned Newton-Krylov implicit MPM method that resamples particles for coarser levels. However, they conclude that such multigrid preconditioning performs similarly to Jacobi preconditioning; this is consistent with our analysis of gemoetric multigrid in Section\ \ref{sec:results}.


On the other hand, Galerkin multigrid\ \cite{strangAndArikka1986} automatically handles boundary conditions by projection. 
However, smooth projection matrices often deteriorate sparsity with large increases in the nonzero entries in coarse level systems.
Xian et al.\ \shortcite{xian2019multigrid} designed their special Galerkin projection criterion based on skinning space coordinates with piecewise constant weights to maintain sparsity, 
but their projection could potentially lead to singular coarser level matrices and thus extra care needs to be taken.
In our work, we derive prolongation and restriction operators via node embedding. Our resulting model is then consistent with an MPM-customized Galerkin multigrid while, due to the regularity of the MPM grid, our resulting coarse level matrices both maintain sparsity via an appropriate choice of kernel and are full-rank.

As in Ferstl et al.\ \shortcite{mcadams2010parallel}, McAdams et al.\ \shortcite{AMGAdaptgrid}, and Zhang et al.\ \shortcite{Zhang2016-hv}, a natural approach would then be to apply our multigrid as a preconditioner in a Krylov solver. 
However, as demonstrated in our benchmark experiments, this straightforward application would not outperform existing diagonally-preconditioned alternatives (PN-PCG) because of the repeated expense of  hierarchy reconstruction at each Newton iterate. 
Instead we develop HOT by applying our multigrid model as an efficient inner initializer within a quasi-Newton solver. 

\subsection{Quasi-Newton Methods}
Quasi-Newton methods e.g. L-BFGS, have long been applied for simulating elastica\ \ \cite{Deuflhard2011-pf}. L-BFGS can be highly effective for minimizing potentials. However, an especially good choice of initializer is required and makes an enormous difference in convergence and efficiency\ \cite{nocedal:2006:numerical}. Directly applying a lagged Hessian at the beginning of each time step is of course the most straightforward option which effectively introduces second order information\ \cite{Brown2013}; unfortunately, it is generally a too costly option with limitations in terms of scalability. Liu et al.\ \shortcite{liu2017quasi} propose to instead invert the Laplacian matrix which approximates the rest-shape Hessian as initializer. This provides better scalability and more efficient evaluations, but convergence speed drops quickly in nonuniform deformation cases\ \cite{Li:2019:DOT}. Most recently Li et al.\ \shortcite{Li:2019:DOT} propose a highly efficient domain-decomposed initializer for mesh-based FE that leverage start of time step Hessians --- providing both scalability and fast convergence in challenging elastodynamic simulations.
For the MPM setting, inexact rather than direct methods are required to approximate the system Hessian given the scale and changing sparsity patterns of MPM simulations. 
Following Li et al.\ \shortcite{Li:2019:DOT}, HOT applies our new multigrid as an inner initializer for L-BFGS to build an efficient method that outperforms or closely matches best-per-example prior methods across all tested cases on state-of-the-art, heavily optimized implicit MPM codes.
Unlike Wen and Goldfarb\ \shortcite{wen2009line} which requires many configuration parameters to alternate between multigrid and single-level solves and uses L-BFGS as the solver for certain multigrid levels,
HOT consistently applies V-cycles on our node embedding multigrid constructed from the projected Hessian\ \cite{teran2005robust} as inner initializer,
without the need of any parameter tuning.





\section{Problem Statement and Preliminaries} \label{sec:background}

\subsection{Optimization-based Implicit MPM}

MPM assembles a hybrid Lagrangian-Eulerian discretization of dynamics. A background Cartesian grid acts as the computational mesh while material states are tracked on particles. 
%
In the following we apply subscripts $p,q$ for particles and $i,j,k$ for grid quantities respectively. We then remove subscripts entirely, as in $\mathbf{\zeta}$, to denote vectors constructed by concatenating nodal quantities $\mathbf{\zeta}_i$ over all grid nodes. Superscripts $n$, and $n+1$ distinguish quantities at time steps $t^n$, and $t^{n+1}$. 
An implicit MPM time step with implicit Euler from $t^n$ to $t^{n+1}$ is performed by applying the following operation sequence:
\begin{description}
\item[Particles-to-grid (P2G) projection.] Particle masses $m_p^{n}$ and velocities $\vv_p^{n}$ are transferred to the grid's nodal masses $m_i^{n}$ and velocities $\vv_i^{n}$ by APIC\ \cite{jiang:2015:apic}.
\item[Grid time step.] Nodal velocity increments, $\Delta \vv_i$, are computed by minimizing implicit Euler's incremental potential in (\ref{eqn:objective}) and are then applied to update nodal velocities by $\vv_i^{n+1}=\vv_i^{n}+\Delta\vv_i$.
\item[Grid-to-particles (G2P) interpolation.] Particle velocities $\vv_p^{n+1}$ are interpolated from $\vv_i^{n+1}$ by APIC.
\item[Particle strain-stress update.] Particle strains (e.g. deformation gradients $\FF_p$) are updated by the velocity gradient $\nabla\vv$ via the updated Lagrangian. Where appropriate, inelasticity is likewise enforced through per-particle strain modification\ \cite{stomakhin:2013:snow,gao2017adaptive}.
\item[Particle advection.] Particle positions are advected by $\vv_p^{n+1}$.
\end{description}
Here we focus on developing an efficient and robust nonlinear solver for the above MPM Grid time step operation. All other operations are standard for MPM (ref. \cite{jiang:2016:course}). 

Assuming an MPM nodal-position-dependent potential energy $\Phi(\xx)$, e.g. a hyperelastic energy, Gast et al. \shortcite{gast:2015:tvcg} observe that minimization of 
\begin{equation} 
E(\Delta\vv) = \sum_i \frac{1}{2}m_i^{n} \|\Delta \vv_i\|^2 + \Phi\left(\xx^{n} + \Delta t (\vv^{n} + \Delta \vv) \right) \label{eqn:objective} 
\end{equation}
subject to proper boundary conditions is equivalent to solving the MPM implicit Euler update $\ff_i(\xx_i^{n}+\Delta t \vv_i^{n+1}) =  (\vv_i^{n+1}-\vv_i^{n})m_i^{n}/\Delta t$, where $\ff_i$ is the implicit nodal force. Minimization of a corresponding incremental potential for the mesh-based elasticity has been widely explored for stable implicit Euler timestepping \cite{bouaziz2014projective,liu2017quasi,Overby2017-ns,Li:2019:DOT}. For MPM, however, a critical difference is that nodal positions $\xx_i$ are virtually displaced from the Eulerian grid during the implicit solve, and are then reset to an empty Cartesian scratchpad. Significantly, across time steps the system matrix can change sparsity pattern. This changing sparsity, together with large MPM system sizes (where more than 100K DOFs are common) generally motivate the application of matrix-free Newton-Krylov methods rather than direct factorization in existing MPM codes.


\begin{figure}[t]
    \centering
    \includegraphics[draft=\mydraft,width=\linewidth]{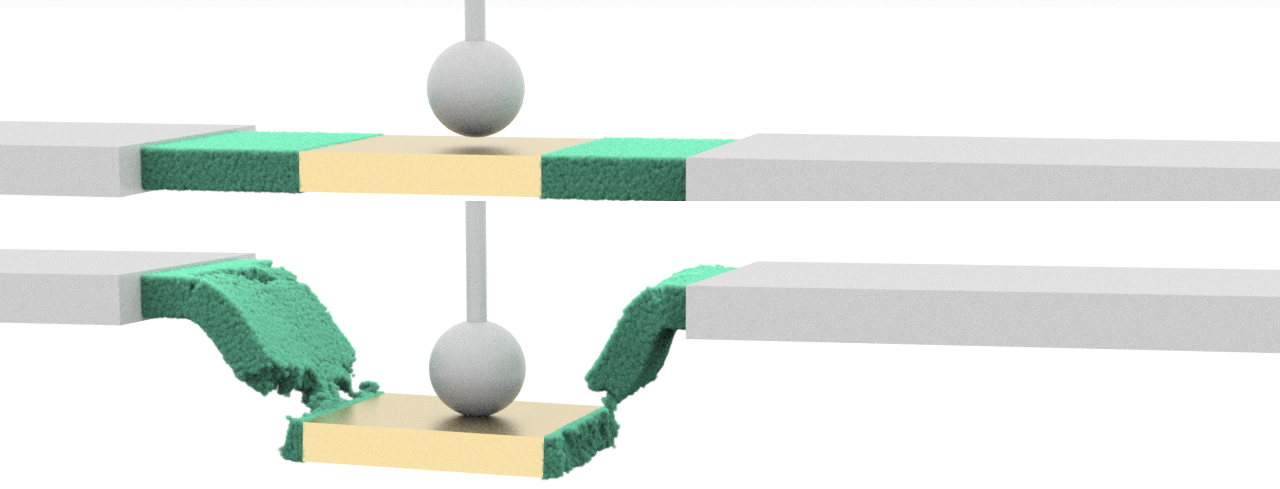}
    \caption{\textbf{Boxes.} A metal box is concatenated with two elastic boxes on both sides. As the sphere keeps pushing the metal box downwards, the elastic boxes end up being torn apart.}
    \label{fig:box}
\end{figure}

\begin{algorithm}[t]
	\caption{Inexact Newton-Krylov Method} \label{alg:InexactPN}
	
	\textbf{Given:} $E$, $\epsilon$ \\
	\textbf{Output:} $\Delta \vv^n$ \\
	
	\textbf{Initialize and Precompute:} \\

	\hspace{20pt}$i \leftarrow 1$, \hspace{3pt} $\Delta \vv^1 \leftarrow \textbf{0}$\\
	
    \hspace{20pt}$g^1 \leftarrow \nabla E(\Delta \vv^1)$ \hspace{5pt} // $E$ is defined in Eq.\ \ref{eqn:objective} \\
	
$\textbf{while}$ $ \text{scaledL2norm}(g^i) > \epsilon \sqrt{n_{\text{node}}}$ $\textbf{do}$ \hspace{5pt} // termination criteria  (\S\ref{sec:CN})\\

\hspace{10pt} $\PP^i \leftarrow \text{projectHessian}(\nabla^2 E(\Delta \vv^i))$ \hspace{5pt} // \cite{teran2005robust} \\

\hspace{10pt} $k \leftarrow \text{min}(0.5, \sqrt{ \text{max}(\sqrt{g_i^{T} P g_i}, \tau)})$  \hspace{5pt} // adaptive inexactness (\S\ref{subsec:inexactcriteria}) \\

\hspace{10pt} $p^i \leftarrow \text{ConjugateGradient}(\PP^i, \textbf{0}, -g^i, k)$ \hspace{5pt} // $k$ as relative tolerance \\

\hspace{10pt} $\alpha \leftarrow \text{LineSearch}(\Delta \vv^i, 1, p^i, E)$  \hspace{5pt} // back-tracking line search \\

\hspace{10pt} $\Delta \vv^{i + 1} \leftarrow \Delta \vv^i + \alpha p^i$ \\

\hspace{10pt} $g^{i + 1} \leftarrow \nabla E(\Delta \vv^{i+1})$ \\

\hspace{10pt}$i \leftarrow i + 1$\\
$\textbf{end while}$ \\

$\Delta \vv^n \leftarrow \Delta \vv^i$
	
\end{algorithm}

\subsection{Inexact Newton-Krylov methods} \label{subsec:inexactnewton}

To minimize (\ref{eqn:objective}) 
with Newton-Krylov methods
further computational savings can be achieved by employing inexact Newton where computational effort in early Newton iterations can be saved by inexactly solving the linear system. For example, Gast et al.\ \shortcite{gast:2015:tvcg} apply the L2 norm of the incremental potential's gradient to adaptively terminate Krylov iterations. 
However, Gast and colleagues mainly target softer materials. However, more generally materials often have large material stiffnesses, e.g., Youngs at $10^9$ for the metal wheel in Fig.~\ref{fig:wheel}.
It becomes even more challenging when materials with widely varying stiffnesses interact with each other. In these cases the inexact Newton strategy in Gast et al.\ \shortcite{gast:2015:tvcg} can simply fail to converge in practical time; e.g., in our experiments for the scenes in Figs.\ \ref{fig:twist} and \ref{fig:chain}.

This observation has motivated the question as to whether an early termination criterion for Newton-type iterations can be computed to obtain visually consistent and stable results across varying simulation inputs. Li et al.\ \shortcite{Li:2019:DOT} extend the characteristic norm (CN) from distortion optimization\ \cite{bcqn18} to elastodynamics and demonstrate its capability to obtain consistent, relative tolerance settings across a wide set of elastic simulation examples over a range of material moduli and mesh resolutions. However, for a scene with materials with drastically different stiffness parameters, the averaging L2 measure will not suffice to capture the multiscale incremental potential gradient in a balanced manner.

We thus propose an extended scaled-CN to support multi-material applications in MPM. Incremental potential gradients are nonuniformly scaled so that multiscale residuals can be effectively resolved. We apply this new characteristic norm to both terminate outer Newton iterations and to improve the inexact Newton iterations in our baseline PN solver. See Algorithm \ref{alg:InexactPN} for our inexact Newton; details are in Section\ \ref{sec:LBFGSMG}.

With extended CN and improved inexact Newton, iterative methods can still suffer from ill-conditioning from stiff materials and so we require preconditioning. Unfortunately incomplete Cholesky is not suitable as elastodynamic system Hessians are not M-matrices\ \cite{kershaw1978incomplete}, leading us to multigrid strategies. However, multigrid construction costs may not be well compensated by the resulting convergence improvement with Newton-Krylov. We thus apply our custom MPM multigrid, next constructed in Section\ \ref{sec:method} below, as an inner initializer inside our quasi-Newton loop; see Section\ \ref{sec:LBFGSMG}.


\begin{figure}[t]
    \centering
    \includegraphics[draft=\mydraft,width=\linewidth]{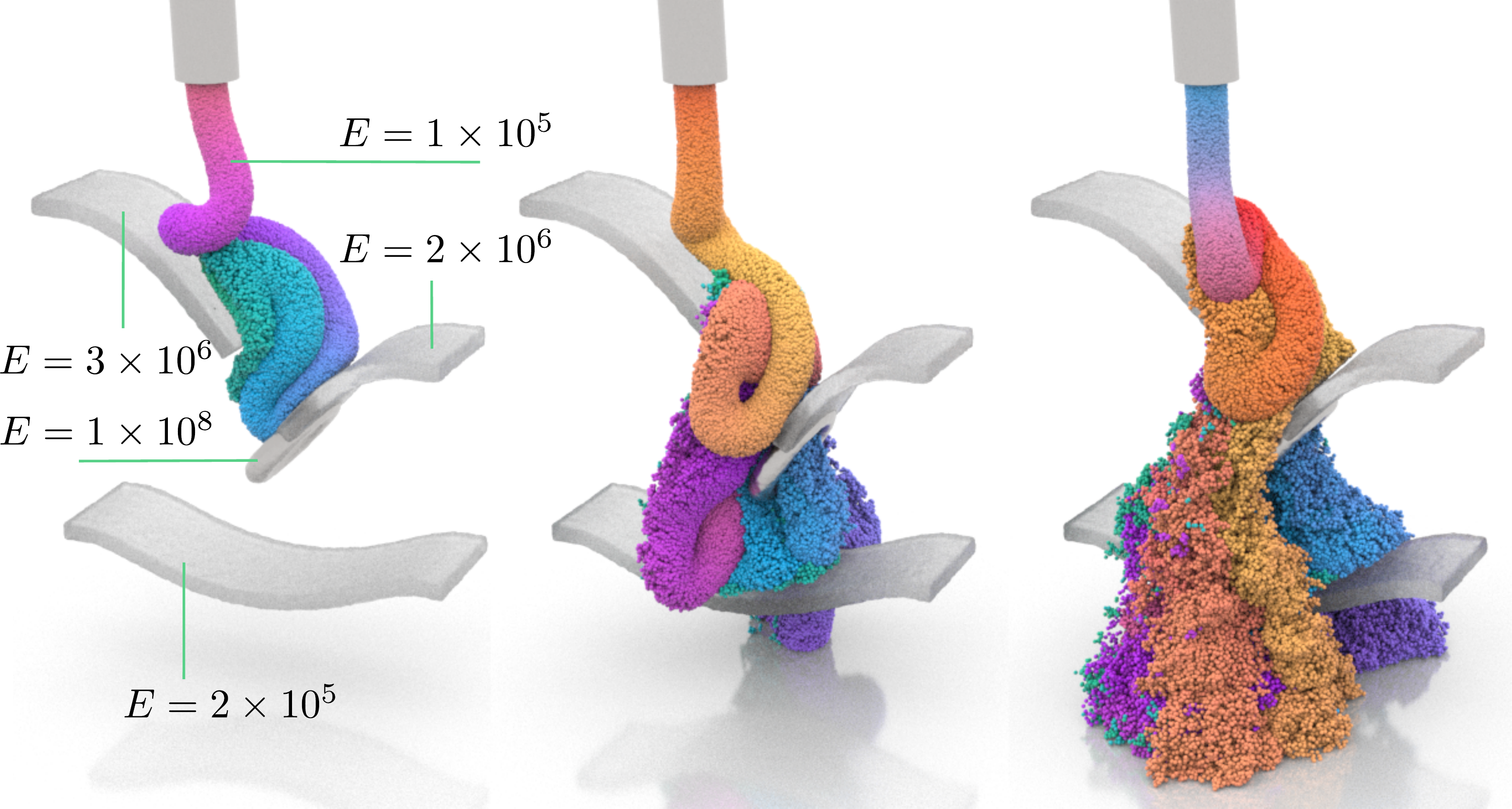}
    \caption{\textbf{Boards.} A granular flow is dropped onto boards with varying Young's moduli, generating coupled dynamics.}
    \label{fig:flow}
\end{figure}
\section{MPM-Multigrid} \label{sec:method}

We propose to construct our hierarchy by embedding finer level grids into the coarser level grids analogously to MPM's embedding of particles into grid nodes.
Then by explicitly storing the system matrix we progressively constructing coarser level matrices directly from the adjacent finer level matrix entries, avoiding the need to compute or store any coarser level geometric information.
We next show that our multigrid is consistent with Galerkin multigrid where boundary conditions are automatically handled and by selecting different node embedding kernels we support flexible control on sparsity.

\subsection{Node-embedding multigrid derivation} \label{subsec:amg}
We begin with an $\mathcal{M}$-level multigrid hierarchy. We denote level $0$ and level $\mathcal{M}-1$ as the finest and coarsest levels respectively. 
System matrices are constructed at each level with prolongation, $\mathcal{P}^{m+1}_m$, and restriction operators, $\mathcal{R}^{m+1}_{m}$, between adjacent levels $m$ and $m+1$. 

We illustrate the construction of our restriction and prolongation operators by considering operations between levels $0$ and $1$. 
Nodal forces in the finest level are
\begin{align}
    \ff_i^0 = - \sum_p V_p \frac{\partial \phi(\xx_p)}{\partial \xx_i^0} 
            = - \sum_p V_p \PP_p \FF_p^T \nabla \omega_{ip}^0.
\end{align}
Here $V_p$ is the initial particle volume, $\phi$ is the energy density function, $\PP_p$ is the first Piola-Kirchhoff stress, $\FF_p$ is the deformation gradient and $\omega_{ip}$ are corresponding particle-grid interaction weights.

In the multigrid hierarchy, residuals, following forces, are restricted from finer to coarser levels.
%
Forces at nodes $j$ in the next level are then
$\ff_j^1 = - \sum_p V_p \frac{\partial \phi(\xx_p)}{\partial \xx_j^1}$.
Embedding finer level nodes to coarser level nodes, we then can simply apply the chain rule, converting derivatives evaluated at a coarse node to those already available at the finer level:
\begin{align}
    \ff_j^1 &= - \sum_i \sum_p V_p  \Big(\frac{\partial \xx_i^0}{\partial \xx_j^1}\Big)^T \frac{\partial \phi(\xx_p)}{\partial \xx_i^0}  
            =  \sum_i \Big(\frac{\partial \xx_i^0}{\partial \xx_j^1}\Big)^T \ff^0_i. \label{eqn:forceoldform}
\end{align}
This gives our restriction operation as $\ff^1 = \mathcal{R}_0^1\ff^0 $ with $\mathcal{R}_0^1 = (\frac{\partial \xx^0}{\partial \xx^1})^T$.

Prolongation is correspondingly given by the transpose $\mathcal{P}_0^1 = (\mathcal{R}_0^1)^T$. 
Recalling that MPM particle velocities $\vv_p$ are interpolated from grid node velocities $\vv_i$ as $\vv_p = \sum_i w_{ip} \vv_i$, 
we have
\begin{equation}
    \begin{aligned}
        \vv_j^0 = \sum_i \frac{\partial \xx^0_i}{\partial \xx^1_j} \vv_i^1 = \sum_i (\mathcal{R}_0^1)_{ji}^T \vv_i^1,
    \end{aligned}
\end{equation}
giving us $\vv^0 = \mathcal{P}_0^1 \vv^1 = (\mathcal{R}_0^1)^T \vv^1$. 

For matrix coarsening we similarly can compute the second-order derivative of (\ref{eqn:objective}) w.r.t. $x^1$. Applying chain rule, with $x^0$ as intermediate variable, we obtain
\begin{equation}
    \begin{aligned}
    (\HH^1)_{jk} = \frac{\partial \ff^1_j}{\partial \xx^1_k} & = \sum_l \frac{\partial \sum_i \Big(\frac{\partial \xx_i^0}{\partial \xx_j^1}\Big)^T \ff^0_i}{\partial \xx^0_l} \frac{\partial \xx^0_l}{\partial \xx^1_k} \\
    & = \sum_i \sum_l \Big(\frac{\partial \xx^0_i}{\partial \xx_j^1}\Big)^T (\HH^0)_{il} \frac{\partial \xx^0_l}{\partial \xx^1_k}.
    \end{aligned}
\end{equation}
Here $\HH^0$ is the Hessian of (\ref{eqn:objective}) w.r.t. $x^0$. We then have the Galerkin operator 
\begin{align}
    \HH^1 = \mathcal{R}_0^1 \HH^0 \mathcal{P}_0^1,
    \label{eq:buildMultigrid}
\end{align}
confirming our construction is consistent with Galerkin multigrid. 
Dirichlet boundary conditions are then resolved at all levels by projection of the corresponding rows and columns of the system matrix and entries in the right-hand-side.



\begin{figure}[b]
    \centering
    \includegraphics[draft=\mydraft,width=\linewidth]{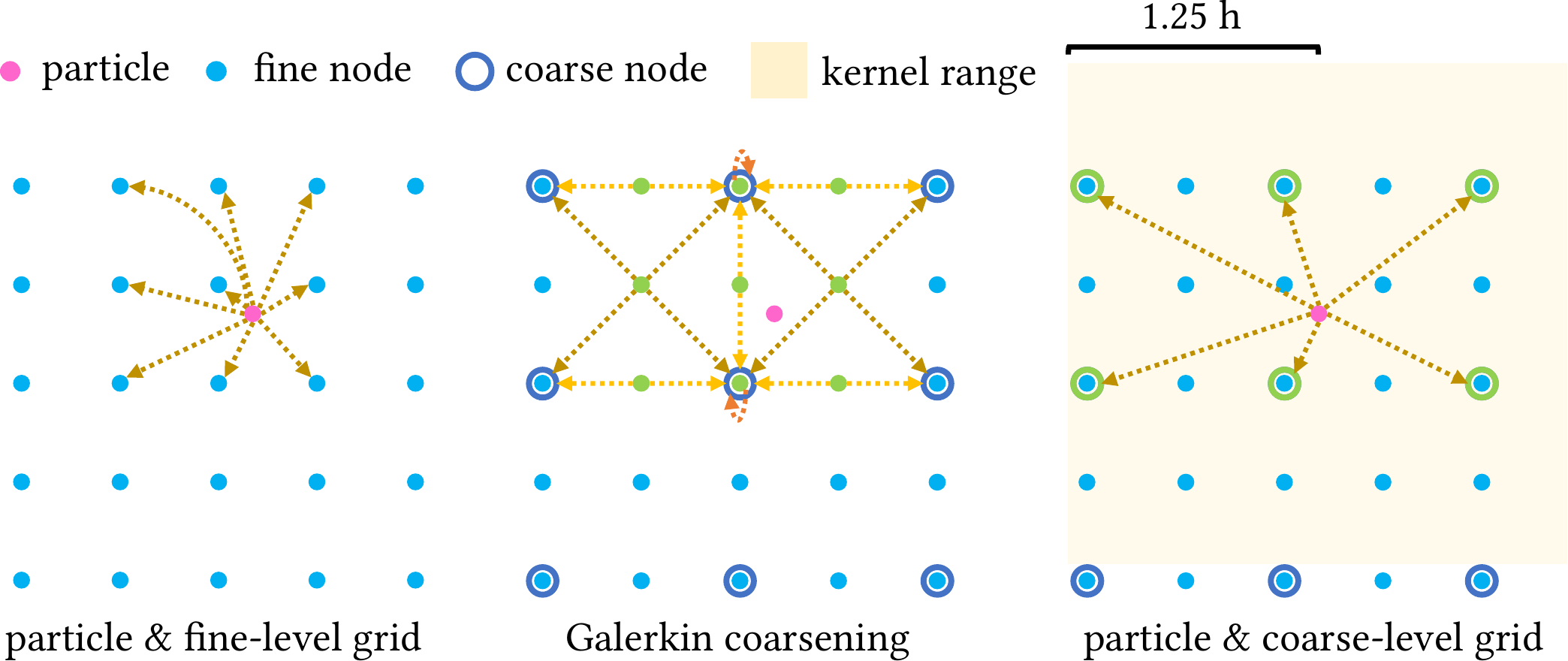}
    \caption{\textbf{Geometric equivalence.} Left: in the finest level, particles' properties are transferred to the grid nodes via the B-spline quadratic weighting function; middle: then the finer nodes transfer information to coarser nodes via linear embedding relationships --- based on which we perform Galerkin coarsening; right: Galerkin coarsening can then be re-interpreted as a new weighting function, with a smaller kernel width, connecting coarser nodes directly with the particles.}
    \label{fig:amggmg}
\end{figure}

\subsection{Geometric multigrid perspective and kernel selection}

Our multigrid is now complete up to our choice of embedding kernel, $\frac{\partial \xx^0}{\partial \xx^1}$, and MPM particle-grid kernel, $\omega_{ip}$. A careful choice of kernels enables us to maintain sparsity as we coarsen. This allows us to improve both convergence and cost.
We apply MPM kernels for our node embedding. Convolution with the particle-grid kernel can then be viewed as a direct embedding of our finest-level \emph{particles} into coarser-level \emph{grids}. This provides a geometric multigrid perspective where we can consider coarse grid matrices as constructed from fine particle quadratures. We next apply this perspective to select our multigrid kernels 

\begin{figure}[t]
    \centering
    \includegraphics[draft=\mydraft,width=0.95\linewidth]{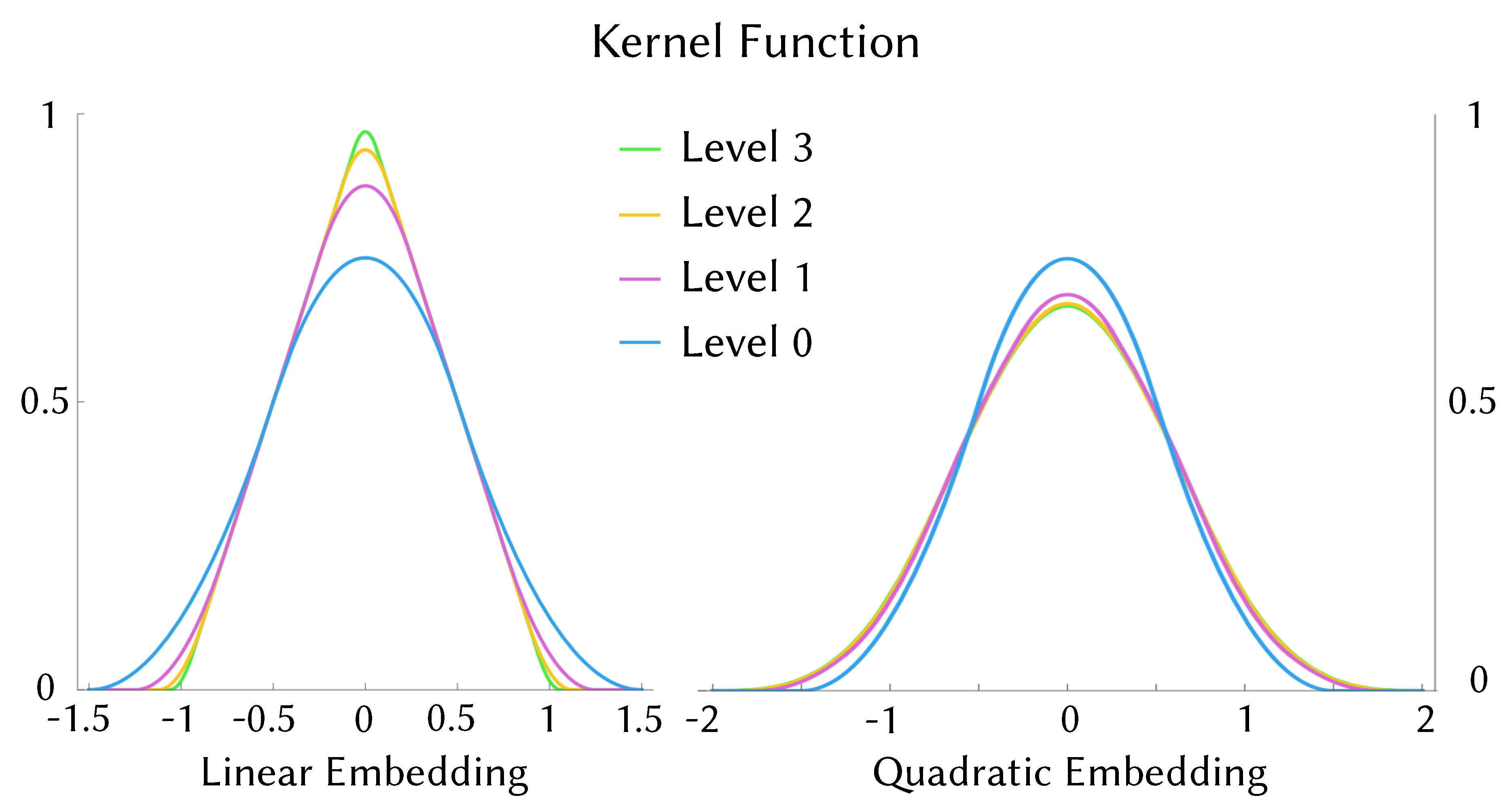}
    \caption{\textbf{Kernel width.} The width of our geometric weighting function, equivalently in our algebraic derivation, changes with level increase. For linear embedding (our choice for HOT), width becomes smaller with coarsening while for quadratic embedding width becomes larger but with an upper bound at 2.}
    \label{fig:our_kernel}
 \end{figure}

We start with a direct MPM derivation defining nodal forces at level $1$:
\begin{align}
    \ff_j^1 = - \sum_p V_p \PP_p \FF_p^T \nabla \omega_{jp}^1.
\end{align}
Compare this with a reformulation of (\ref{eqn:forceoldform}) where we apply our newly defined restriction operator:
\begin{align}
    \ff_j^1 = - \sum_p V_p \PP_p \FF_p^T ( \sum_i (\mathcal{R}_0^1)_{ji}\nabla \omega_{ip}^0 ).
\end{align}
Here particle-grid weight gradients between level $1$ and particles are now given by $\sum_i (\mathcal{R}_0^1)_{ji}\nabla \omega_{ip}^0$, and our multigrid obtains a simple geometric interpretation as illustrated in Fig. \ref{fig:amggmg}.  
As a geometric multigrid, this provides a weighting function directly bridging between particles and coarse grid nodes. The grid itself can be generated by traversing all particles to find occupied coarse nodes. Similarly, a concatenation of prolongation operators for each coarse level, right-multiplied by the original weight gradient, gives us the new weight gradients required in each successive level. In turn, with this weight gradient, the Hessian matrix can be defined to complete the geometric multigrid model. We use the corresponding weighting function to plot curves in Fig.~\ref{fig:our_kernel}.

\begin{figure}[t]
    \centering
    \includegraphics[draft=\mydraft,width=0.9\linewidth]{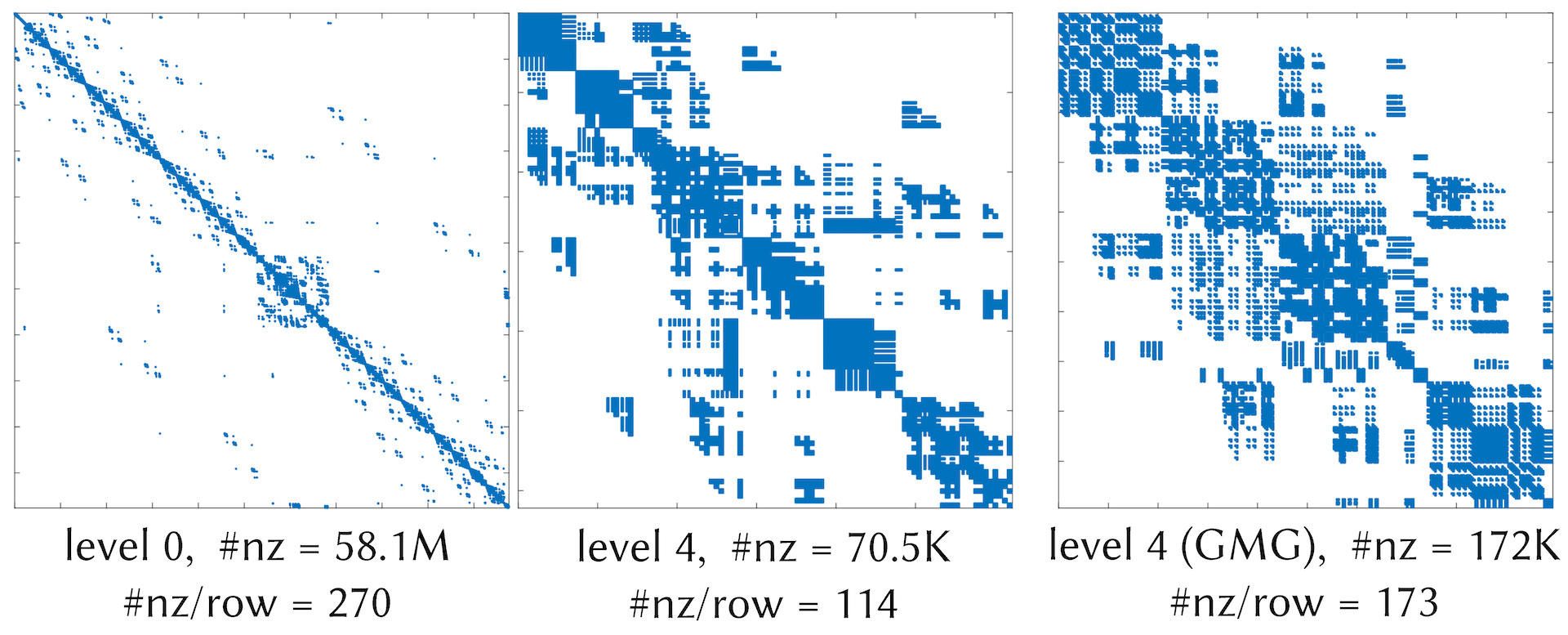}
    \caption{\textbf{Sparsity pattern.} Our MPM multigrid system matrices gain better sparsity as levels increase because stencil sizes decrease. Left and middle: our level-0 and level-4 matrices for the ArmaCat simulation in Fig.~\ref{fig:cat}. Right: direct geometric multigrid generates denser matrices for increasingly coarse levels; here shown at  level $4$ for the same simulation. Note that although in simulation we only used a 3-level multigrid, we here illustrate the sparsity patterns with 5 levels for visual clarity.}
    \label{fig:sparsityPattern}
 \end{figure}

For HOT we apply B-spline quadratic weighting for our base particle-grid kernel and choose the linear kernel for our embedding. The latter defines our prolongation and restriction operators between adjacent levels in the hierarchy.
With this choice the stencil size of our coarser level systems become progressively smaller, providing better sparsity.
As shown in Fig.~\ref{fig:our_kernel} left, kernel width reduces from $3\Delta x$ to $2\Delta x$ as levels increase.
An alternative would be to uniformly apply the B-spline \emph{quadratic} weighting for all kernels. However, stencil size would then grow as we coarsen (c.f. Fig.~\ref{fig:our_kernel} right) making it computationally less attractive; see Table 2 in our supplemental document for the comparison. Likewise, direct geometric multigrid, where particles are directly coarsened, also exhibits impractical fill-in as stencil sizes grow with coarsening. See for example Fig.~\ref{fig:sparsityPattern} where we compare the matrix sparsity patterns for the ArmaCat simulation in Fig.~\ref{fig:cat}.

\section{Hierarchical Optimization Time Integration} \label{sec:LBFGSMG}

Our newly constructed MPM multigrid structure can be used as a preconditioner by applying one V-cycle (Algorithm\ \ref{alg:vcycle}) per iteration for a conjugate gradient solver to achieve superior convergence in a positive-definite fixed, inexact Newton method. In the following we denote this approach as ``projected Newton, multigrid preconditioned conjugate gradient'' or PN-MGPCG. 
However, in practice, the cost of reconstructing the multigrid hierarchy at each Newton iteration 
of PN-MGPCG is not well-compensated by the convergence improvement,  providing only little or moderate speedup compared to a baseline projected Newton PCG solver (PN-PCG) (see Figs.\ \ref{fig:CN} and\ \ref{fig:wedge}) where a simple diagonal preconditioner is applied to CG.
This is because in PN-MGPCG, each outer iteration (PN) requires reconstructing the multigrid matrices, and each inner iteration (CG) performs one V-cycle.
One reconstruction of the multigrid matrices would take around $4 \times$ the time for one V-cycle and over $20\times$ the time for one Jacobi preconditioned PCG iteration.
Unlike the Poisson system in Eulerian fluids simulation, the stiffnesses of elastodynamic systems are often not predictable as it varies a lot under different time step sizes, deformation, and dynamics.
Therefore, it is hard for PN-MGPCG to consistently well accelerate performance in all time steps.

\subsection{Multigrid initialized quasi-Newton method} \label{subsec:HOT}

Rather than applying MPM multigrid as a \emph{preconditioner} for a Krylov method (which can still both be slow and increasingly expensive as we grow stiffness; see Fig.\ \ref{fig:wedge}), inspired by Li et al.\ \shortcite{Li:2019:DOT}, we apply our MPM multigrid as an \emph{inner initializer} for a modified L-BFGS solver. In the resulting hierarchical method, multigrid then provides efficient second-order information, while our outer quasi-Newton low-rank updates\ \cite{Li:2019:DOT} provide efficient curvature updates to maintain consistent performance for time steps with widely varying stiffness, deformations and conditions.
In turn, following recent developments we choose a start of time step lagged model update \cite{Brown2013,Li:2019:DOT}. We re-construct our multgrid structure once at the start of each time step solve. This enables local second order information to efficiently bootstrap curvature updates from the successive light-weight, low-rank quasi-Newton iterations. 
  

\begin{algorithm}[h!]
	\label{alg:vcycle}
	\caption{Multigrid V-Cycle Preconditioner}
	\textbf{Given:} $\mathcal{R}$, $\mathcal{P}$, $\mathcal{M}$ \hspace{10pt} \\
	\textbf{Input:} $b^0$, $\HH$ \\
	\textbf{Output:} $u^0$ \\
	\vspace{5pt}
	\hspace{0pt}\textbf{for} $m = 0, 1,..,\mathcal{M}-2$ \\
    \hspace{10pt}$u^m \leftarrow \textbf{0}$ \\
	\hspace{10pt}$u^m \leftarrow \text{SymmetricGaussSeidel}({\HH^m}, u^m, b^m)$ \\
	\hspace{10pt}$b^{m+1} \leftarrow \mathcal{R}_m^{m+1} (b^m - \HH^m u^m)$ \\
    \hspace{0pt}\textbf{end for}

    \hspace{0pt}$u^{\mathcal{M}-1} \leftarrow \text{ConjugateGradient}({\HH^{\mathcal{M}-1}}, u^{\mathcal{M}-1}, b^{\mathcal{M}-1}, 0.5)$ \\

	\hspace{0pt}\textbf{for} $m = \mathcal{M}-2, \mathcal{M}-3,..,0$ \\
    \hspace{10pt}$u^m \leftarrow u^m + \mathcal{P}_m^{m+1} u^{m+1}$ \\
	\hspace{10pt}$u^m \leftarrow \text{SymmetricGaussSeidel}({\HH^m}, u^m, b^m)$ \\
    \hspace{0pt}\textbf{end for}
    \label{alg:vcycle}
\end{algorithm}

This completes the core specifications of our Hierarchical Optimization Time (HOT) integrator algorithm.
The HOT multigrid hierarchy is constructed at the beginning of each time step.
Then, for each L-BFGS iteration, the multiplication of our initial Hessian inverse approximation to the vector is applied by our multigrid V-cycle.
To ensure the symmetric positive definiteness of the V-cycle operator, we apply colored symmetric Gauss-Seidel as the smoother for finer levels and employ Jacobi preconditioned CG solves for our coarsest level system (see Algorithm\ \ref{alg:vcycle}). We apply PCG for our coarsest level rather than a direct solve as the subtle convergence improvement overhead of could not compensate for the overhead of factorization; see Section\ \ref{subsec:sgs_jacobi}. While weighted Jacobi is effectively applied in Eulerian fluid simulation\ \cite{Zhang2016-hv} as a smoother for multigrid, here, in testing, we observe that determining proper weighting that obtains efficient or even convergent behavior for non-diagonally dominant elastodynamic Hessians is challenging. Similarly, we found Chebyshev smoothers\ \cite{adams2003parallel} impractical as estimating reasonable upper and lower eigenvalues of the system matrix introduces unacceptably large overhead.

\begin{figure}[t]
    \centering
    \includegraphics[width=\linewidth]{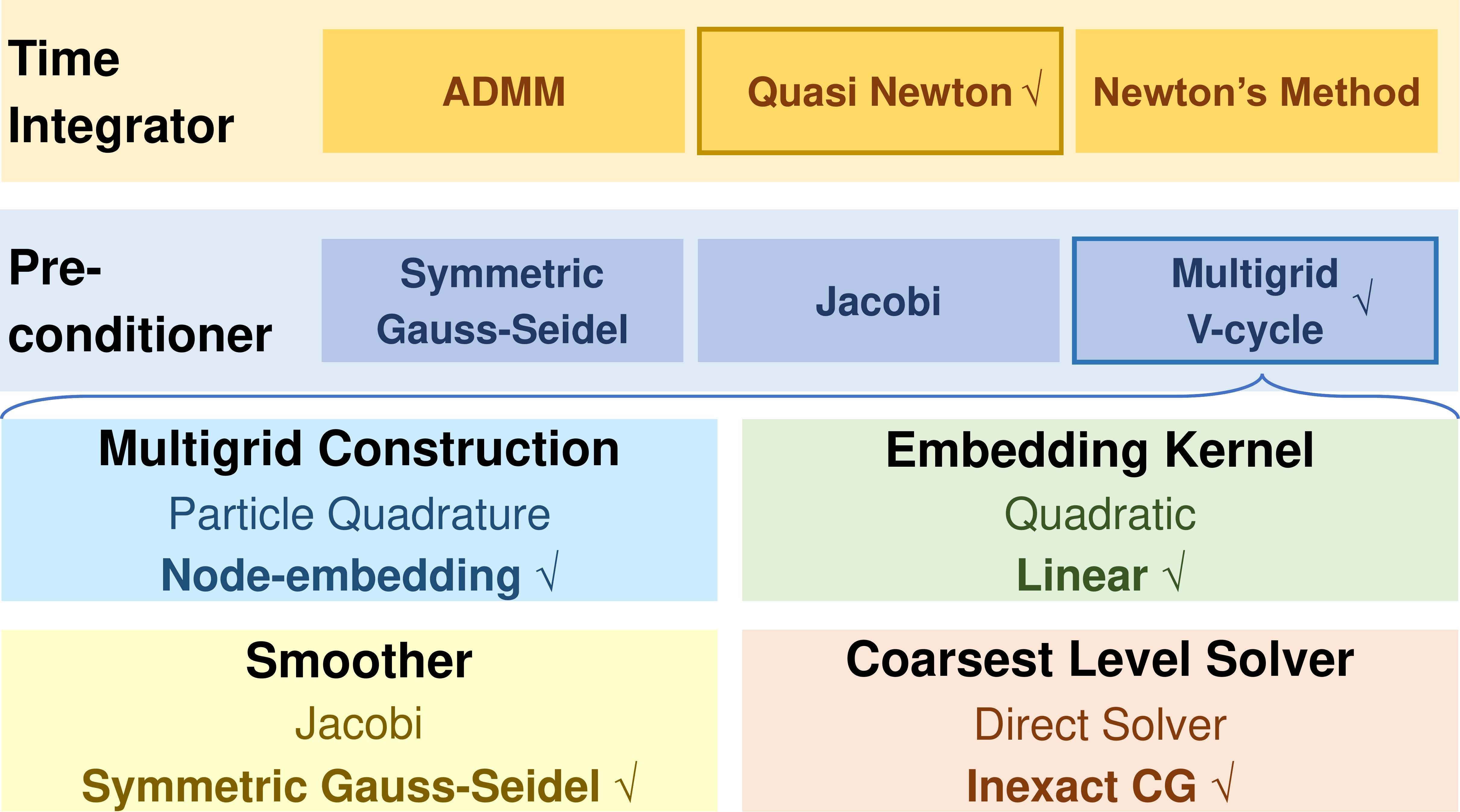}
    \caption{\textbf{Summary of HOT's Design Choices.} }
    \label{fig:designChoices}
\end{figure}

HOT's curvature information is updated by low-rank secant updates with window size $w=8$, producing a new descent direction for line search at each L-BFGS iteration.
Pseudocode for the HOT method is presented in Algorithm~\ref{alg:HOT}. We analyze its performance, consistency and robustness in Sec. \ref{sec:performance} with comparisons to state-of-the-art MPM solvers. In Figure\ \ref{fig:designChoices} we highlight design choices for HOT together with superficially reasonable alternatives that we compare and analyze in Section\ \ref{sec:results}.

\begin{algorithm}[t]
	\caption{Hierarchical Optimization Time Integrator (HOT)} \label{alg:HOT}
	
    \textbf{Given:} $E$, $\epsilon$, $w$, $\mathcal{R}$, $\mathcal{P}$  \hspace{10pt}
    
    \textbf{Output:} $\Delta \vv^{n}$
	
	\textbf{Initialize and Precompute:}
	
    \hspace{20pt}$i \leftarrow 1$, \hspace{3pt} $\Delta \vv^1 \leftarrow \textbf{0}$\\

    \hspace{20pt}$g^1 \leftarrow \nabla E(\Delta \vv^1)$ \hspace{5pt} // $E$ is defined in Eq.\ \ref{eqn:objective}\\
    
    \hspace{20pt}$\PP^1 \leftarrow \text{projectHessian}(\nabla^2 E(\Delta \vv^1))$ \hspace{5pt} // \cite{teran2005robust}
    
    \hspace{20pt}$\HH \leftarrow \text{buildMultigrid}(\PP^1, \mathcal{R}, \mathcal{P})$ \hspace{5pt} // Eq.\ \ref{eq:buildMultigrid}\\
    
    \vspace{3pt}

	//  Quasi-Newton loop to solve time step $n+1$: \\
	$\textbf{while}$ $ \text{scaledL2norm}(g^i) > \epsilon \sqrt{n_{\text{node}}}$ $\textbf{do}$\hspace{5pt} // termination criteria  (\S\ref{sec:CN})\\
	
\hspace{10pt} $q\leftarrow -g^i$ 

\vspace{2pt}
\hspace{10pt} // L-BFGS low-rank update \\
\hspace{10pt} \textbf{for} $a = i-1, i-2,..,i-w$ \hspace{5pt} // break if $a < 1$

\hspace{20pt} $s^a \leftarrow \Delta \vv^{a+1} - \Delta \vv^a, \> \> \> y^a \leftarrow g^{a+1} - g^a,  \> \> \> \rho^a  \leftarrow 1/((y^a)^Ts^a)$

\hspace{20pt} $\alpha^a \leftarrow \rho^a (s^a)^T q $ 

\hspace{20pt} $q \leftarrow q - \alpha^a y^a$ 


\hspace{10pt} \textbf{end for}

\vspace{2pt}
\hspace{10pt} $r \leftarrow \text{V-cycle}(q, \HH)$ \hspace{5pt} // Algorithm\ \ref{alg:vcycle}


\vspace{2pt}
\hspace{10pt} // L-BFGS low-rank update \\
\hspace{10pt} \textbf{for} $a = i-w, i-w+1,..,i-1$ \hspace{5pt} // skip (continue) until $a \geq 1$

\hspace{20pt} $\beta \leftarrow \rho^a (y^a)^T r$ 

\hspace{20pt} $r \leftarrow r + (\alpha^a - \beta) s^a$

\hspace{10pt} \textbf{end for}

\hspace{10pt} $p^i \leftarrow r$

\hspace{10pt} $\alpha \leftarrow \text{LineSearch}(\Delta \vv^i, 1, p^i, E)$  \hspace{5pt} // back-tracking line search 

\hspace{10pt} $\Delta \vv^{i+1}  \leftarrow  \Delta \vv^i + \alpha p^i$  

\hspace{10pt} $g^{i+1} \leftarrow \nabla E(\Delta \vv^{i+1})$

\hspace{10pt} $i \leftarrow i+1$

$\textbf{end while}$

$\Delta \vv^n \leftarrow \Delta \vv^i$
	
\end{algorithm}

\subsection{Convergence tolerance}\label{sec:CN}
To reliably obtain consistent results across heterogenous simulations while saving computational effort,
we extend the Characteristic Norm (CN)\ \cite{bcqn18, Li:2019:DOT} from FE mesh to MPM discretization, taking multi-material domains into account.
To simulate multiple materials with significantly varying material properties, coupled in a single simulated system, 
the traditional \emph{averaging} L2 measure fails to characterize the multiscale incremental potential's gradient. 

For MPM we thus first derive a node-wise CN in MPM discretization, 
and then set tolerances with the L2 measure of the node-wise CN scaled incremental potential gradient.
Concretely, we compute the norm of the stress derivative evaluated at the undeformed configuration in the diagonal space, $\xi_p = ||\frac{d \hat{\PP}}{d \hat{\FF}}||_p$, for each particle $p$, and transfer this scalar field with mass weighting and normalization to corresponding grid node quantities $\xi_i$. 
Here $\xi_i$ is in units of $J/m^3$ as $\PP$ is in the unit of energy density and $\FF$ is unitless.
We then compute a node-wise CN as
\begin{equation}
    \ell_i \xi_i \Delta t,
    \label{eq:nodewiseCN}
\end{equation}
per node where $\ell_i$ characterizes discretization, $\xi_i$ characterizes averaged material stiffness per node, and $\Delta t$ provides time step scaling.
In mesh-based FE, $\ell_i$ is the area of the polyhedron formed by the one-ring elements connecting to node $i$\ \cite{bcqn18}.
For MPM, we correspondingly have 
$\ell_i = 24 \Delta x^2$ from the uniform Cartesian grid discretization. 

To check convergence we scale each entry of the incremental potential gradient vector $\gg$ (in units of $\text{kg}\cdot m/s$ as our optimization variable is velocity) with the corresponding node-wise CN computed in Eq.\ (\ref{eq:nodewiseCN}), obtaining the unitless $\hat{\gg}$. 
Termination queries then compare $||\hat{\gg}||$ against $\epsilon \sqrt{n}$, 
where $n$ gives the number of active grid nodes and $\epsilon$ is the selected accuracy tolerance. 
Note, we confirm that when a single, uniform material is applied in the simulation, our extended CN measure correctly reduces to Li et al.'s\ \shortcite{Li:2019:DOT} L2 measure.

\begin{figure}[t]
    \centering
    \includegraphics[draft=\mydraft,width=\linewidth]{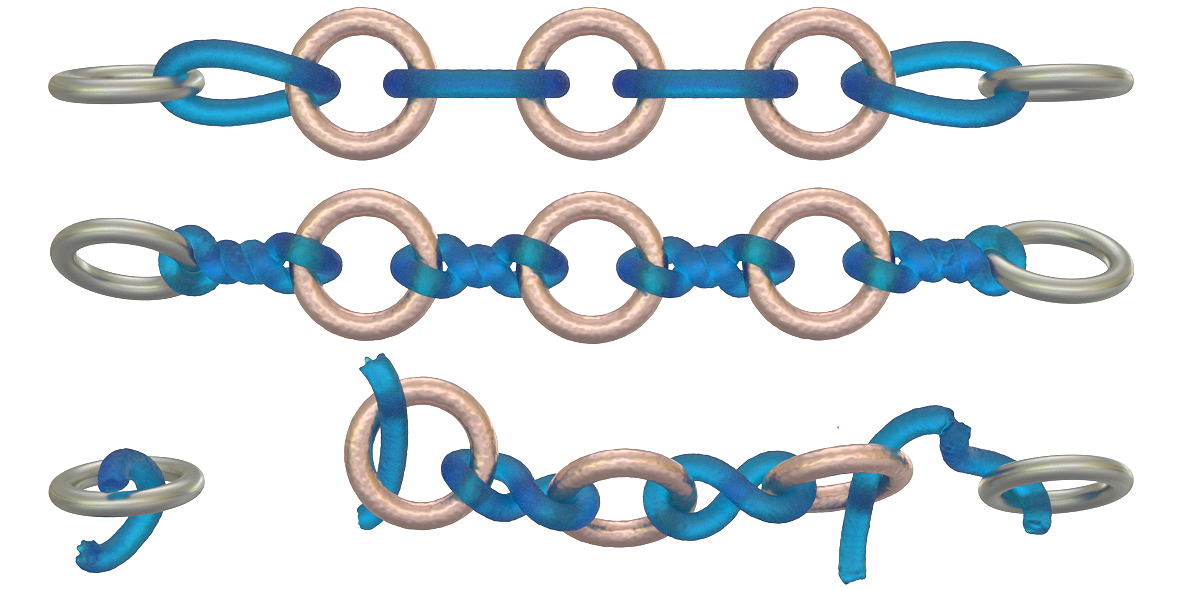}
    \caption{\textbf{Rotating chain.} A chain of alternating soft and stiff rings is rotated until soft rings fracture, dynamically releasing the chain.}
    \label{fig:chain}
\end{figure}

\subsection{Inexact linear solves} \label{subsec:inexactcriteria}
We apply the above extended CN criterion to terminate outer nonlinear solve iterates. Within each outer nonlinear iteration, inner iterations are performed to solve the corresponding linear system to a reasonable accuracy.
For the first few Newton iterations of an inexact Newton-Krylov solve, the initial nonlinear residuals are generally far from convergence and so inexact linear solves are preferred.
As discussed in Gast et al. \shortcite{gast:2015:tvcg}, inexact linear solves can significantly relieve the computational burden in large time step simulations.
Thus they set a relative tolerance $k$ on the energy norm $\sqrt{r_0^T \PP r_0}$, where $\PP$ is the preconditioning matrix, of the initial residual $r_0 = -\nabla E(\Delta \vv^i)$ for each linear solve in Newton iteration $i$. Gast and colleagues use $k = \min(0.5, \sqrt{\max(||\nabla E(\Delta \vv^i)||, \tau)})$, where $\tau$ is a nonlinear tolerance,
and perform CG iterations until the current $\sqrt{r^T \PP r}$ is smaller than $k\sqrt{r_0^T \PP r_0}$.
This strategy can be traced back to classical optimization strategies where setting $k = \min(0.5, \sqrt{||\nabla E(\Delta \vv^i)||})$ can be shown to yield super-linear convergence\ \cite{nocedal:2006:numerical} .

However, this approach is challenged when it comes to heterogeneous materials in two aspects.
First, the L2 norm of the incremental potential does not take into account its multi-scale nature, potentially providing too small relative tolerances for stiff materials.
Second, as discussed above, the nonlinear tolerance in Gast et al.\ \shortcite{gast:2015:tvcg} is challenging to tune per example, especially with stiff material models.
Therefore, we modify this inexact strategy for our baseline PN-PCG by applying $k = \text{min}(0.5, \sqrt{ \text{max}(\sqrt{r_0^{T} P r_0}, \tau)})$ as the relative tolerance to terminate CG iterations.
Here the preconditioning matrix $\PP$ in the energy norm $\sqrt{r_0^T \PP r_0}$ has the effect of locally scaling per-node residuals to account for varying material stiffnesses,
while $\tau$ is simply Li et al.'s\ \shortcite{Li:2019:DOT} tolerance on L2 measure characterizing the most stiff material in the running scene, ensuring that our tolerance will not be too small for stiffer materials.


HOT similarly exploits our inexact solving criterion for the coarsest level PCG during early L-BFGS iterations.
Specifically, in each V-cycle, we recursively restrict the right-hand side vector $\bb_0$ to the coarsest level to obtain $\bb_{m-1}$.  
We then set the tolerance for the CG solver to $1/2\sqrt{\bb_{m-1}^T \DD_{m-1}^{-1} \bb_{m-1}}$ where $\DD_{m-1}$ is the diagonal matrix extracted from the system matrix at level $m-1$. Note that the same V-cycle and termination criterion is also adopted in our PN-MGPCG.
As L-BFGS iterations proceed, the norm of $\bb_{m-1}$ decreases, leading to increasingly accurate solves at the coarsest level.
As demonstrated in Sec.~\ref{sec:results}, this reduces computational effort --- especially when the system matrices at the coarsest level are not well conditioned.

\section{Implementation} \label{sec:imple}
Accompanying this paper, we open source all of our code including scripts for running all presented examples with HOT and all other implemented methods compared with in test code. ADMM MPM\ \cite{Fang:2019:ViscousMPM}, is separately available\ \footnote{\url{https://github.com/squarefk/ziran2019}}.
Here we provide remarks on the nontrivial implementation details that can significantly influence performance.

\paragraph{Lock-free multithreading.}
For all particle-to-grid transfer operations (including each matrix-vector multiplication in PN-PCG(MF)), we adopt the highly-optimized lock-free multithreading from Fang et al. \shortcite{fang2018temporally}. This also enables the parallelization of our colored symmetric Gauss Seidel smoother for the multigrid V-cycle. All optimizations are thus consistently utilized (wherever applicable) across all compared methods so that our timing comparisons more reliably reflect algorithmic advantages of HOT.

\paragraph{Sparse matrix storage.}
We apply the quadratic B-spline weighting kernel for particle fine-grid transfers. 
The number of non-zero entries per row of the system matrix at the finest level can then be up to $5^d$ where \textit{d} denotes dimension. 
In more coarsened levels, the number of non-zero entries decreases due to the linear embedding of nodes in our MPM multigrid, as can be seen from Fig.~\ref{fig:our_kernel}.
Similarly, for our restriction/prolongation matrix, the number of non-zero entries per row/column is $3^d$ for linear kernel.
Notice that in all cases, the maximum number of nonzeros per row can be pre-determined, thus we employ \textit{diagonal storage format} in our implementation to store all three matrix types for accelerating matrix computations. 

\paragraph{Multigrid application}
In our experiments (Section\ \ref{sec:results}), our MPM multigrid is tested both as a preconditioner for the CG solver in each PN-MGPCG outer iteration, and as the inner initializer for each L-BFGS iteration of HOT.

\paragraph{Prolongation and restriction.}
Our prolongation operator is defined as in traditional particle-grid transfers in hybrid methods -- finer nodes are temporarily regarded as particles in the coarser level.
Spatial hashing is then applied to record the embedding relation between finer and coarser grid nodes for efficiency.

\begin{figure}[b]
    \centering
    \includegraphics[draft=\mydraft,width=\linewidth]{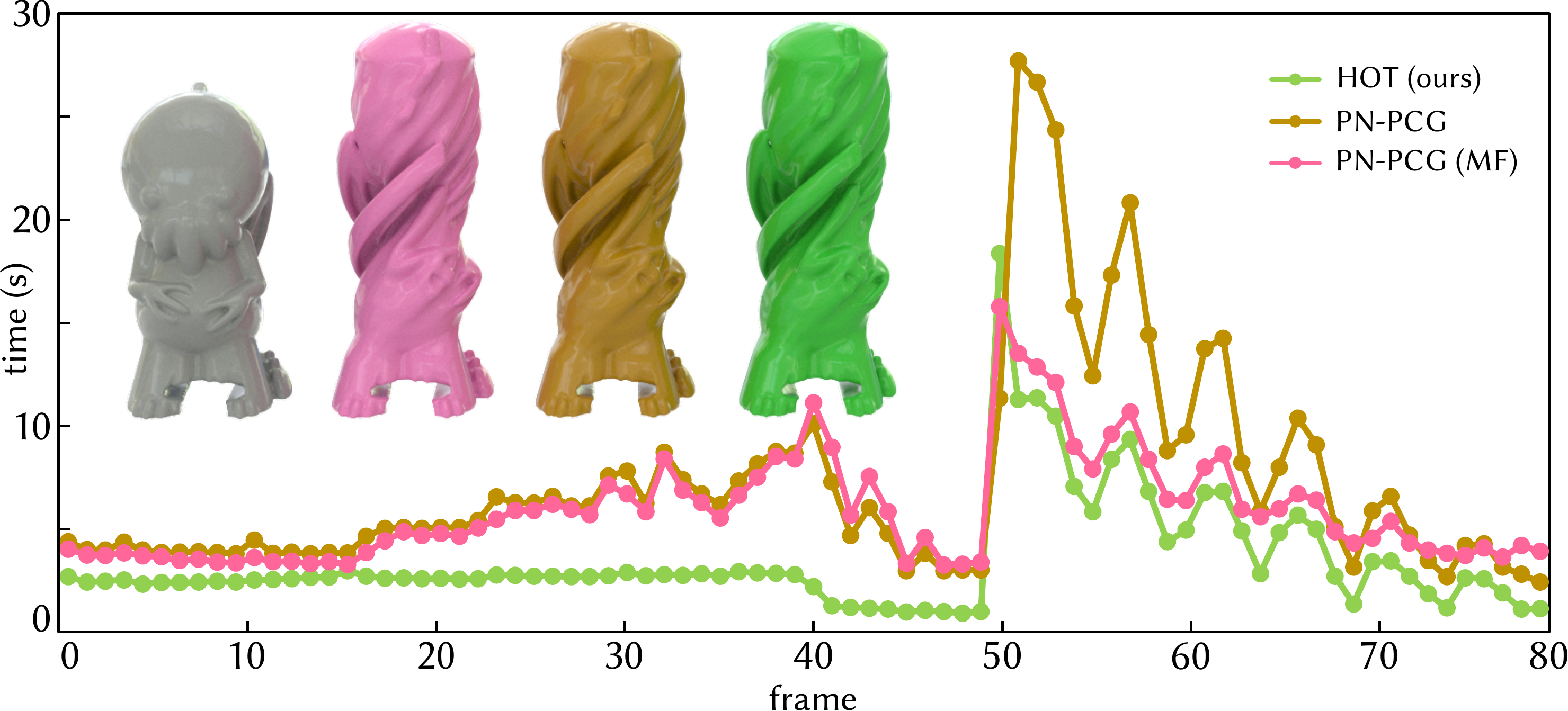}
    \caption{\textbf{Faceless.} We rotate the cap and then release the head of the ``faceless'' mesh. Upon release dynamic rotation and expansion follow. Here we plot the computation time of each single time step frame for HOT and PN-PCG (both matrix-based and matrix-free). HOT outperforms across all time step solves during the simulation.}
    \label{fig:faceless}
\end{figure}
\section{Results and Evaluation} \label{sec:results}

\subsection{Benchmark}
\paragraph{Methods in comparison}
We implement a common test-harness code to enable the consistent evaluation comparisons between HOT and other possible Newton-Krylov methods, i.e. PN-PCG, matrix free (MF) PN-PCG, and state-of-the-art (MF) from Gast et al. \shortcite{gast:2015:tvcg}.
To ensure consistency, PN-PCG and MF PN-PCG adopt our node-wise CN with the same tolerances. For Gast et al.\ \shortcite{gast:2015:tvcg} where a global tolerance for the residual norm is applied, we manually select the largest tolerance value ($10^{-3}$) that produces artifact-free results for all experiments. 
In addition to Gast et al.\ \shortcite{gast:2015:tvcg}, we also compare to the ADMM-based state-of-the-art implicit MPM method\ \cite{Fang:2019:ViscousMPM} on the faceless example to demonstrate  differences in the order of convergence (Figure\ \ref{fig:admm}).
We note that other than Gast et al.\ \shortcite{gast:2015:tvcg} and Fang et al.\ \shortcite{Fang:2019:ViscousMPM}, all other methods in our study are applied here, to our knowledge, for the first time for MPM.

We continue our ablation study here on how our design choices for HOT impact performance and convergence. We compare HOT with other potential new MPM solvers that one may consider designing, including 
(1) HOT-quadratic: HOT's framework with the quadratic (rather than linear) embedding kernel;
(2) LBFGS-GMG: L-BFGS with a more standard geometric multigrid as the initializer;
(3) PN-MGPCG: A Newton-Krylov solver replacing PN-PCG's Jacobi-precoditioned CG with HOT's multigrid-preconditioned CG;
(4) and an MPM extension of the quasi-Newton LBFGS-H (FEM) from Li et al.\ \shortcite{Li:2019:DOT}.
Note that unlike in Li et al.\ \shortcite{Li:2019:DOT} where the LBFGS-H is based on fully factorizing the beginning of time step Hessian with a direct solver, here we only partially invert the hessian by conducting Jacobi preconditioned CG iterations with adaptive terminating criteria identical to that of the coarsest-level solve in HOT (Section \ref{subsec:inexactcriteria}). In other words, it is an inexact LBFGS-H equivalent to a single-level HOT. We find that this inexact LBFGS-H often leads to better performance than those with direct solvers in large-scale problems.

All methods in our ablation study together with Gast et al.\ \shortcite{gast:2015:tvcg} are implemented in C++ and consistently optimized (see Section \ref{sec:imple}). Assembly and evaluations are parallelized with Intel TBB. 

\begin{figure*}[t]
    \centering
    \includegraphics[draft=\mydraft,width=\linewidth]{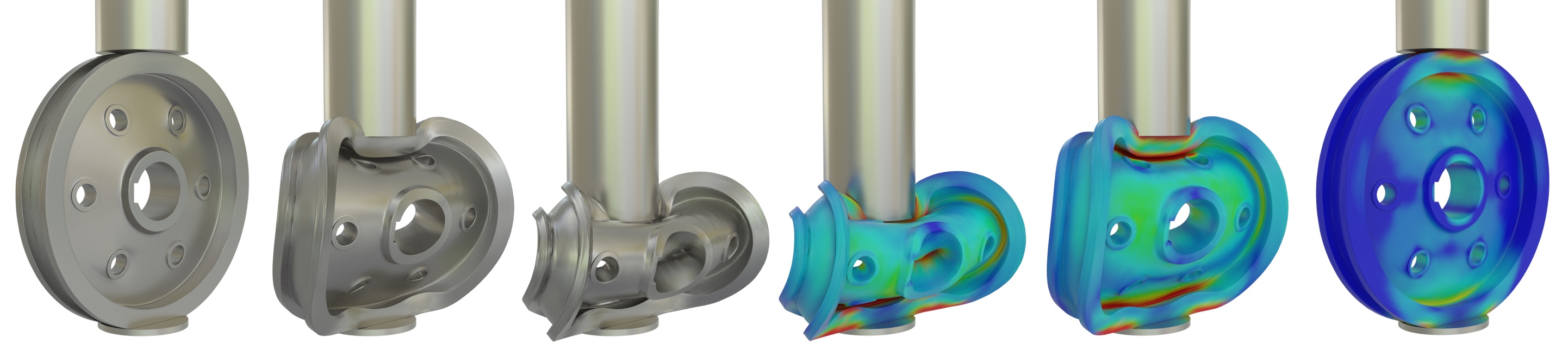}
    \caption{\textbf{Wheel.} Our HOT integrator enables unified, consistent and predictive simulations of metals with real-world mechanical parameters (with stress magnitude visualized). }
    \label{fig:wheel}
\end{figure*}


\begin{figure}[b]
    \centering
    \includegraphics[draft=\mydraft,width=0.95\linewidth]{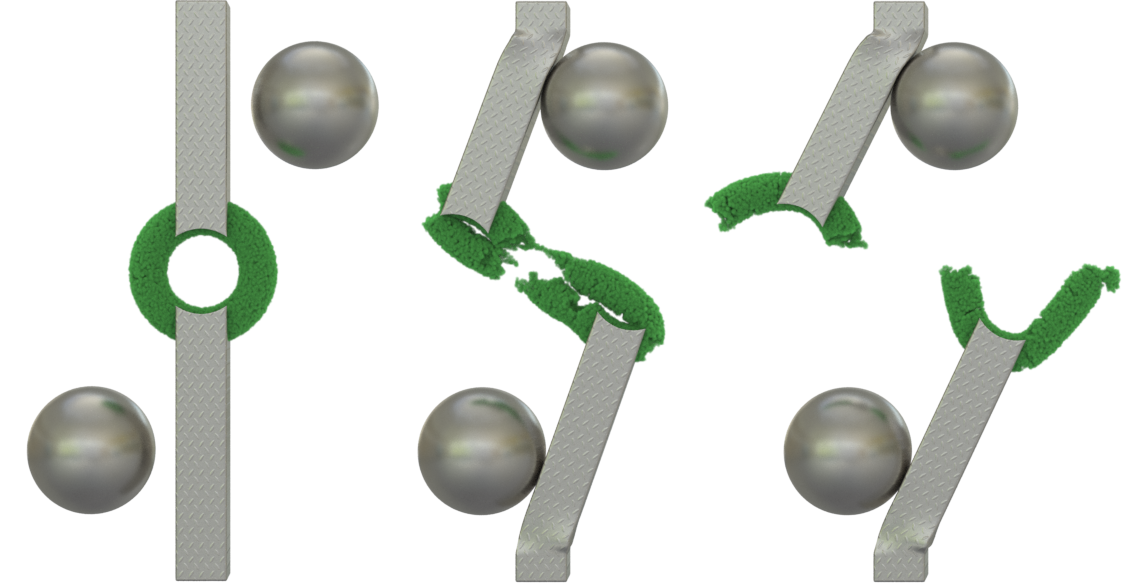}
    \caption{\textbf{Donut.} An elastic torus is mounted between two metal bars. Collisions with rigid balls deform the attaching bars that then break the torus. }
    \label{fig:donut}
\end{figure}
\begin{figure}[t]
    \centering
    \includegraphics[draft=\mydraft,width=\linewidth]{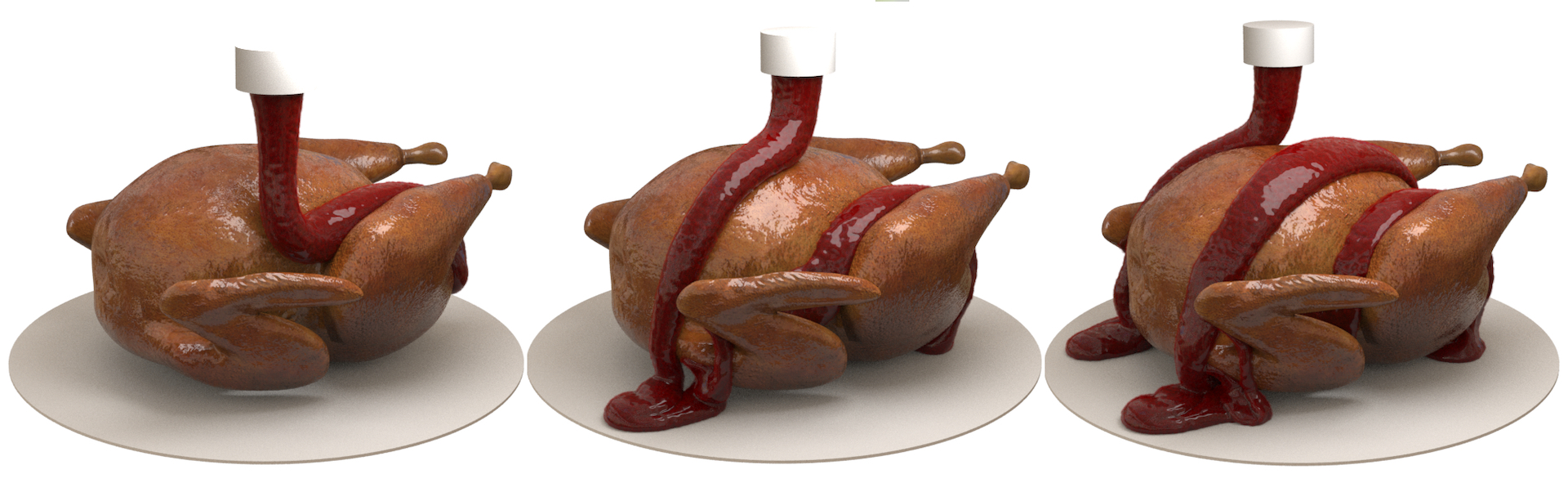}
    \caption{\textbf{HOT Sauce.} HOT sauce is poured onto a turkey.}
    \label{fig:sauce}
\end{figure}

\paragraph{Simulation settings}
Time step sizes are set to $\min(\frac{1}{\text{FPS}}, 0.6 \frac{\Delta x}{v_{\text{max}}})$ throughout all the simulations. Here $0.6$ is selected to fulfill the CFL condition. We observe in our tests that a 3-level multigrid preconditioner with one symmetric Gauss-Seidel smoothing and an inexact Jacobi PCG as our coarsest-level solver works best for both HOT and PN-MGPCG. We observe a window size of 8 for LBFGS methods yields favorable overall performance. In our experiments, across a wide range of scenes, $\epsilon = 10^{-7}$ delivers consistent visual quality for all examples even when we vary materials with widely changing stiffness, shapes and deformations.

Fixed corotated elasticity (FCR) from Stomakhin et al.\ \shortcite{stomakhin:2012:invertible} is applied as our elasticity energy in all examples. 
In addition to our scenes with purely elastic materials: Twist (Fig.\ \ref{fig:twist}), ArmaCat (Fig.\ \ref{fig:cat}), Chain (Fig.\ \ref{fig:chain}), and Faceless (Fig.\ \ref{fig:faceless}), we also test with
plastic models: von Mises for Sauce (Fig. \ref{fig:sauce}) and for metal in Wheel (Fig. \ref{fig:wheel}), the center box in Boxes (Fig.\ \ref{fig:box}), and the bars in Donut (Fig.\ \ref{fig:donut}); and granular flow\ \cite{stomakhin:2013:snow} in Boards (Fig.\ \ref{fig:flow}).

We perform and analyze extensive benchmark studies on challenging simulation settings, where the material parameters are all from real world data, and most of them are heterogenous. 
This not only significantly simplifies the material tuning procedure in animation production, but also helps to achieve realistic physical behavior and intricate coupling between materials with varying stiffness. Fig.\ \ref{fig:stiffness} demonstrates that simulating a scene with aluminum sheets with inappropriate material parameters could end up getting unexpected behavior. 
See Table\ \ref{table:real-materials} for the physical parameters in daily life and Table \ref{table:material} for material parameters used in our benchmark examples.

Detailed timing statistics for all examples are assembled in Table 1 and Table 2 from the supplemental document. 
As discussed in Section\ \ref{sec:CN}, all evaluated methods in our ablation study are terminated with the same tolerance computed from our extended characteristic norm for fair comparisons across all examples.

\subsection{Performance}\label{sec:performance}

\begin{table}[b]
\centering
\fontsize{.62em}{.1em}\selectfont
\caption{\textbf{Material parameters.} The von Mises yield stress: $^\dagger$$2.4\times10^8$ Pa; $^\ddagger$$720$ Pa. $^\star$The plastic flow from Stomakhin et al. \shortcite{stomakhin:2013:snow}, where singular values of the deformation gradient are clamped into $[0.99,1.001]$.}
\setlength{\tabcolsep}{1pt}
\begin{tabular}{l@{\hskip1pt}lrrrrrrrrr}
\toprule
\multicolumn{2}{l}{Example}     & Particle \#   & $\Delta x$ (m)            & density (kg/m$^3$)                                             & Young's modulus (Pa)                  &$\nu$      \\\midrule
(Fig.~\ref{fig:twist}) & Twist & $230$k     & $1\times 10^{-2}$    & $2\times10^3$                                      & $5\times10^5$/$5\times10^9$                   & 0.4   \\\midrule
(Fig.~\ref{fig:box}) & Boxes$^\dagger$       & $805$k    & $8\times 10^{-3}$    & $2\times10^3$/$2.7\times10^3$                      & $2\times10^5$/$6.9\times10^{10}$              & 0.33          \\\midrule
(Fig.~\ref{fig:donut}) & Donut$^\dagger$       & $247$k    & $5.7\times 10^{-3}$  & $2\times10^3$/$2.7\times10^3$                      & $1\times10^5$/$6.9\times10^{10}$              & 0.33          \\\midrule
(Fig.~\ref{fig:cat}) & ArmaCat  & $403$k     & $4\times 10^{-2}$    & $1\times10^3$/$1\times10^3$/$2.5\times10^3$        & $1\times10^5$/$1\times10^6$/$1\times10^{9}$   & 0.4/0.47/0.4   \\\midrule
(Fig.~\ref{fig:chain}) & Chain       & $308$k     & $1\times 10^{-2}$    & $2\times10^3$/$2\times10^3$                        & $5\times10^5$/$3\times10^9$                   & 0.4          \\\midrule
(Fig.~\ref{fig:flow}) & Boards$^\star$      & $188$k    & $7\times 10^{-3}$    & $1\times10^3$                                      & $1\times10^5$-$1\times10^8$                   & 0.33         \\\midrule
(Fig.~\ref{fig:wheel}) & Wheel$^\dagger$       & $550$k     & $5\times 10^{-3}$    & $2.7\times10^3$/$2.7\times10^3$                    & $1\times10^5$/$6.9\times10^{10}$              & 0.4/0.33  \\\midrule
(Fig.~\ref{fig:faceless}) & Faceless    & $110$k      & $1\times 10^{-2}$    & $2\times10^3$                                      & $5\times10^4$                                 & 0.3      \\\midrule
(Fig.~\ref{fig:sauce}) & Sauce$^\ddagger$ & $311$k     & $1.5\times 10^{-2}$    & $2.7\times10^3$                                      & $2.1\times10^5$                   & 0.33   \\\midrule
\end{tabular}

\label{table:material}
\end{table}

We analyze relative performance of HOT in two sets of comparisons. First, we compare HOT against the existing state-of-the-art implicit MPM methods\ \cite{gast:2015:tvcg, Fang:2019:ViscousMPM}. Second, we perform an extensive ablation study to highlight the effectiveness of each of our design choices.

\paragraph{Comparison to state-of-the-art}
As discussed earlier, the performance and quality of results from Gast et al.\ \shortcite{gast:2015:tvcg} are highly dependent, per-example, on the choice of tolerance parameter. Across many simulation runs we sought optimal settings for this parameter to efficiently produce visually plausible, high-quality outputs. However, we find that suitable nonlinear tolerances vary extensively with different simulation conditions such as materials and boundary conditions.
For example, we found an ideal tolerance for the Wheel example (Figure\ \ref{fig:wheel}) at $10^{2}$, while for the Faceless example (Figure\ \ref{fig:faceless}) $10^{-3}$ worked best. On the other hand applying the $10^{2}$ tolerance generates instabilities and even explosions for all other examples (see the supplemental document Figure 1), while using $10^{-3}$ tolerance produces extremely slow performance especially for examples containing stiff materials (see the supplemental document Table 1).
As for ADMM MPM\ \cite{Fang:2019:ViscousMPM}, as it is a first-order method we observe slow convergence. Thus we postpone detailed analysis to our convergence discussion below (Section\ \ref{sec:convergence}).
In contrast HOT requires no parameter tuning. All results, timings and animations presented here and in the following were generated without parameter tuning using the same input settings to the solver. As demonstrated they efficiently  converge to generate consistent visual quality output. 

%

\paragraph{Ablation Study}
We start with the homogeneous ``faceless'' example with a soft elastic material ($E=5\times 10^4$ Pa); we rotate and raise its head and then release. As shown in Fig.\ \ref{fig:faceless}, for this scene with moderate system conditioning, HOT already outperforms the two PN-PCG methods from our ablation set 
in almost every frame. Here there is already a nearly 2$\times$ average speedup of HOT for the full simulation sequence compared to both the two PN-PCG variations; while the overall maximum speedup per frame is around 6$\times$.

We then script a set of increasingly challenging stress-test scenarios across a wide range of material properties, object shapes, and resolutions; see, e.g., Figs.\ \ref{fig:cat} and\ \ref{fig:sauce} as well as the supplemental video. For each simulation we apply HOT with three levels 
so that the number of nodes is generally several thousand or smaller at the coarsest level. In Table 1 from the supplemental document we summarize runtime statistics for these examples comparing HOT's total wall clock speedup for the entire animation sequence, and maximum speedup factor per frame compared to PN-PCG, PN-PCG(MF), PN-MGPCG, and LBFGS-H across the full set of these examples.

\begin{figure}[t]
    \centering
    \includegraphics[draft=\mydraft,width=\linewidth]{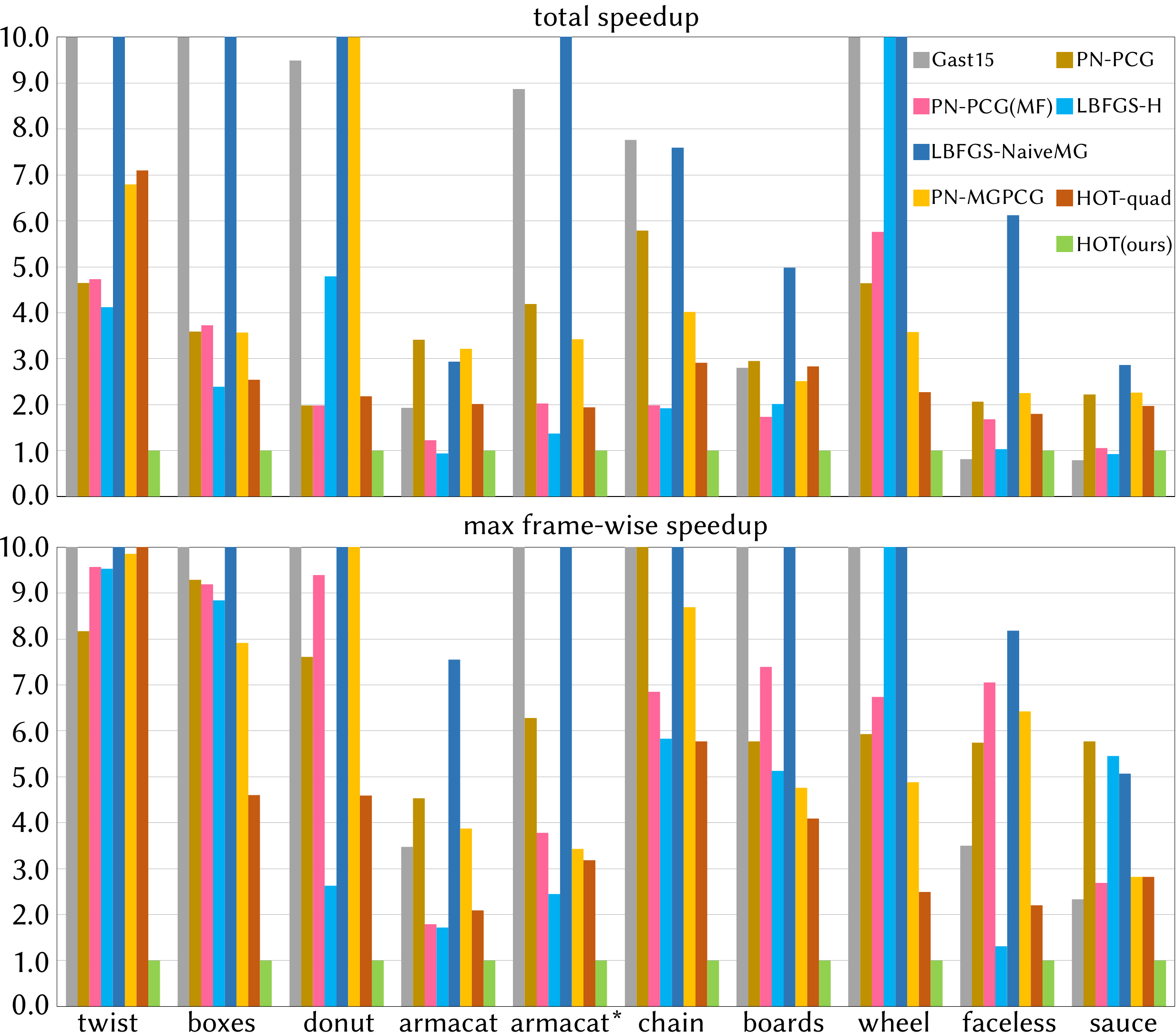}
    \caption{\textbf{Speedup overview.} Top: we summarize method timings for all benchmark examples measuring the total runtime of each method normalized w.r.t the timing of the HOT algorithm (``how HOT'') over each simulation sequence; and so determine HOT's speed-up. Bottom: we comparably report the normalized maximum frame-wise timing of each method w.r.t. HOT across all benchmark examples and so again determine per-frame max speed-up of HOT. Here each simulation example is labeled at bottom, where the cat in armacat and armacat* is with $1MPa$ and $1GPa$ Young's modulus respectively.}
    \label{fig:speedup}
\end{figure}

\paragraph{Timings}
Across this benchmark set we observe HOT has the fastest runtimes, for all but two examples (see below for discussion of these), over the best timing for each example across all methods: PN-PCG, PN-PCG(MF), PN-MGPCG, and LBFGS-H. Note that these variations for our ablation study already well exceed the state-of-the-art method from Gast et al. in most examples. In general HOT ranges from 1.98$\times$ to 5.79$\times$ faster than PN-PCG, from 1.05$\times$ to 5.76$\times$ faster than PN-PCG(MF), from 2.26$\times$ to 10.67$\times$ faster than PN-MGPCG, and from 1.03$\times$ to 4.79$\times$ faster than LBFGS-H on total timing. The exceptions we observe are for the easy Sauce (Young's $2.1\times10^5$Pa) and ArmaCat (Young's $10^6$Pa) examples, where materials are already quite soft and the deformation is moderate. In these two simple examples HOT performs closely to the leading LBFGS-H method.
However, when simulations become challenging we observe that LBFGS-H can have trouble converging. This is most evident in the stiff aluminum Wheel example (Fig.\ \ref{fig:wheel}), where the metal is undergoing severe deformation. Here HOT stays consistently efficient, outperforming all other methods. See our convergence discussion below for more details.
Importantly, across examples, we observe that alternate methods PN-PCG, PN-PCG(MF), PN-MGPCG, and LBFGS-H swap as fastest per example so that it is never clear which would be best a priori as we change simulation conditions.
While in some examples each method can have comparable performance within 2$\times$ slower than HOT, they also easily fall well behind to both HOT and other methods in other examples (Fig.\ \ref{fig:speedup}). In other words, none of these other methods have even empirically consistent good performance across tested examples. The seemingly second best LBFGS-H can even fail in some extreme cases.
For most of the scenes with heterogenous materials or large deformations, e.g. Twist, Boxes, Donut, and Wheel, which results in more PN iterations, PN-PCG is faster than its matrix-free counterpart PN-PCG(MF). Among these examples only Boxes and Wheel can be well-accelerated by using MGPCG for PN.

\paragraph{Gauss-Seidel preconditioned CG} \label{subsec:sgs_jacobi}
Here we additionally compare the symmetric Gauss-Seidel (SGS) and Jacobi preconditioned PN-PCG to show that a simple trade-off between convergence and per-iteration cost might not easily lead to significant performance gain (Table \ref{tb:sgs_jacobi}).
SGS preconditioned PN-PCG significantly improves convergence compared to Jacobi preconditioning as one would expected, but due to its more expensive preconditioning computation, the performance is right at the same level and sometimes even worse.
This is also why we applied Jacobi preconditioned CG to solve our coarsest level system.
\begin{table}[t]
    \caption{\textbf{Jacobi v.s. Gauss-Seidel preconditioned PN-PCG:} Here we compare symmetric Gauss-Seidel and Jacobi preconditioned PN-PCG. The runtime environment for all benchmark examples are identical to Table 1 from the supplemental document. \textit{avg time} measures average absolute cost (seconds) per playback frame, \textit{avg iter} measures the average number of PCG iterations (per method) required per time step to achieve the requested accuracy.}
      \setlength{\tabcolsep}{1pt}
    \small\setlength{\tabcolsep}{.2em}
      \begin{tabular}{rrrrrrrrrrrrrrr}
          \toprule
          \multicolumn{1}{c|}{\multirow{2}{*}{Example}}  & \multicolumn{2}{c|}{SGS}                                       & \multicolumn{2}{c}{Jacobi}\\
          \multicolumn{1}{c|}{}                          & \multicolumn{1}{c}{avg time} & \multicolumn{1}{c|}{avg CG iter}   & \multicolumn{1}{c}{avg time} & \multicolumn{1}{c}{avg CG iter} \\ \hline
          Twist                                          & 492.20                    & 679.91                              & 361.53             & 1054.16 \\
          Boxes                                          & 1368.54                    & 34.13                               & 466.94             & 248.98 \\
          Donut                                          & 410.34                    & 45.59                              & 240.65             & 375.39 \\
          ArmaCat (soft)                              & 109.08                     & 29.04                               & 111.04             & 66.43 \\
          ArmaCat (stiff)                             & 148.38                     & 62.63                               & 153.84             & 157.77 \\
          Chain                                          & 398.79                     & 47.27                               & 572.01             & 92.67 \\
          Boards                                         & 371.66                    & 59.17                               & 313.62             & 206.35 \\ 
          Wheel                                          & 269.08                     & 106.13                               & 206.14             & 424.13 \\
          Faceless                                       & 12.22                      & 16.14                               & 7.21             & 63.80 \\
          Sauce                                          & 22.66                     & 9.98                               & 29.07             & 27.06 \\\bottomrule
      \end{tabular} 
      \label{tb:sgs_jacobi}
  \end{table}

\paragraph{Changing machines}
Across different consumer-level Intel CPU machines we tested (see the supplemental document Table 1), we see that HOT similarly maintains the fastest runtimes across all machines regardless of available CPU or memory resources, over the best timing for each example between PN-PCG, PN-PCG(MF), PN-MGPCG, and LBFGS-H.

\subsection{Convergence} \label{sec:convergence}
HOT balances efficient, hierarchical updates with global curvature information from gradient history. In this section we first compare HOT's convergence to the state-of-the-art ADMM MPM\ \cite{Fang:2019:ViscousMPM}, and then analyze the convergence behavior based on our ablation study. Here we exclude Gast et al. as the method applies a different stopping criteria and, as discussed above, requires intensive parameter tuning.

\begin{figure}[t]
    \centering
    \includegraphics[draft=\mydraft,width=\linewidth]{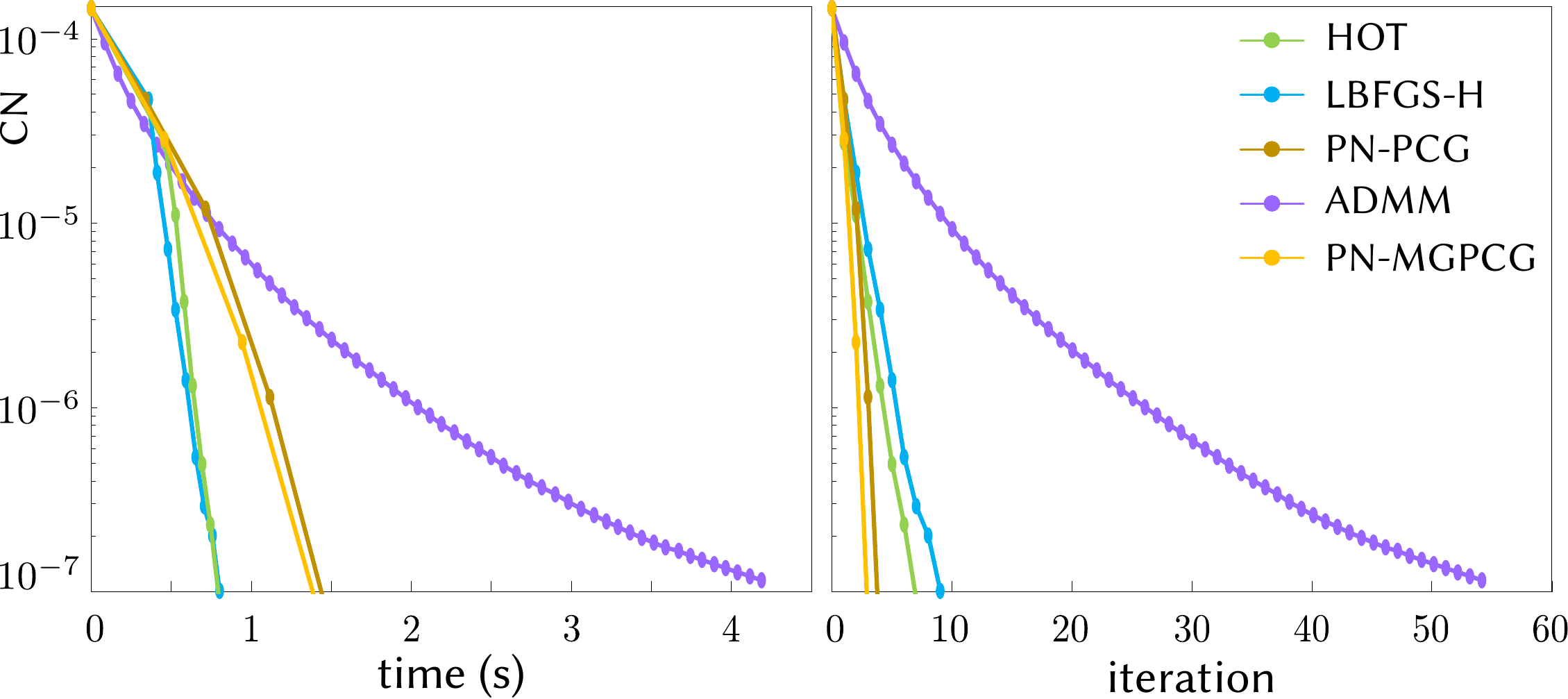}
    \caption{\textbf{Comparison to ADMM-MPM.} CN-scaled gradient norm to timing and iteration plots for the first time step of the faceless example (Fig.\ \ref{fig:faceless}) of all methods including ADMM-MPM\ \cite{Fang:2019:ViscousMPM}.
    With much lower per-iteration cost, ADMM quickly reduces the residual within first few iterations (left). However, as a first-order method it converges slowly to the requested accuracy compared to all others which converges super-linearly. As a result, it is $20 \times$ slower than our HOT. }
    \label{fig:admm}
\end{figure}

\paragraph{Comparison to ADMM}
Here we compare to the ADMM-MPM\ \cite{Fang:2019:ViscousMPM} on a pure elasticity example faceless (Fig.\ \ref{fig:faceless}) by importing their open-sourced code into our codebase and adopted our nodewise CN based termination criteria (Section\ \ref{sec:CN}).
Despite their advantages on efficiently resolving fully-implicit visco-elasto-plasticity, on this pure elasticity example we observe that as a first-order method, ADMM converges much slower than all other Newton-type or Quasi-Newton methods including HOT (Figure\ \ref{fig:admm}).
Although the overhead per iteration of ADMM is generally few times smaller, the number of iterations required to reach the requested accuracy is orders of magnitudes more. Nevertheless, ADMM-MPM is more likely to robustly generate visually plausible results within first few iterations, while Newton-type or Quasi-Newton methods might not.

\begin{figure}[t]
    \centering
    \includegraphics[draft=\mydraft,width=\linewidth]{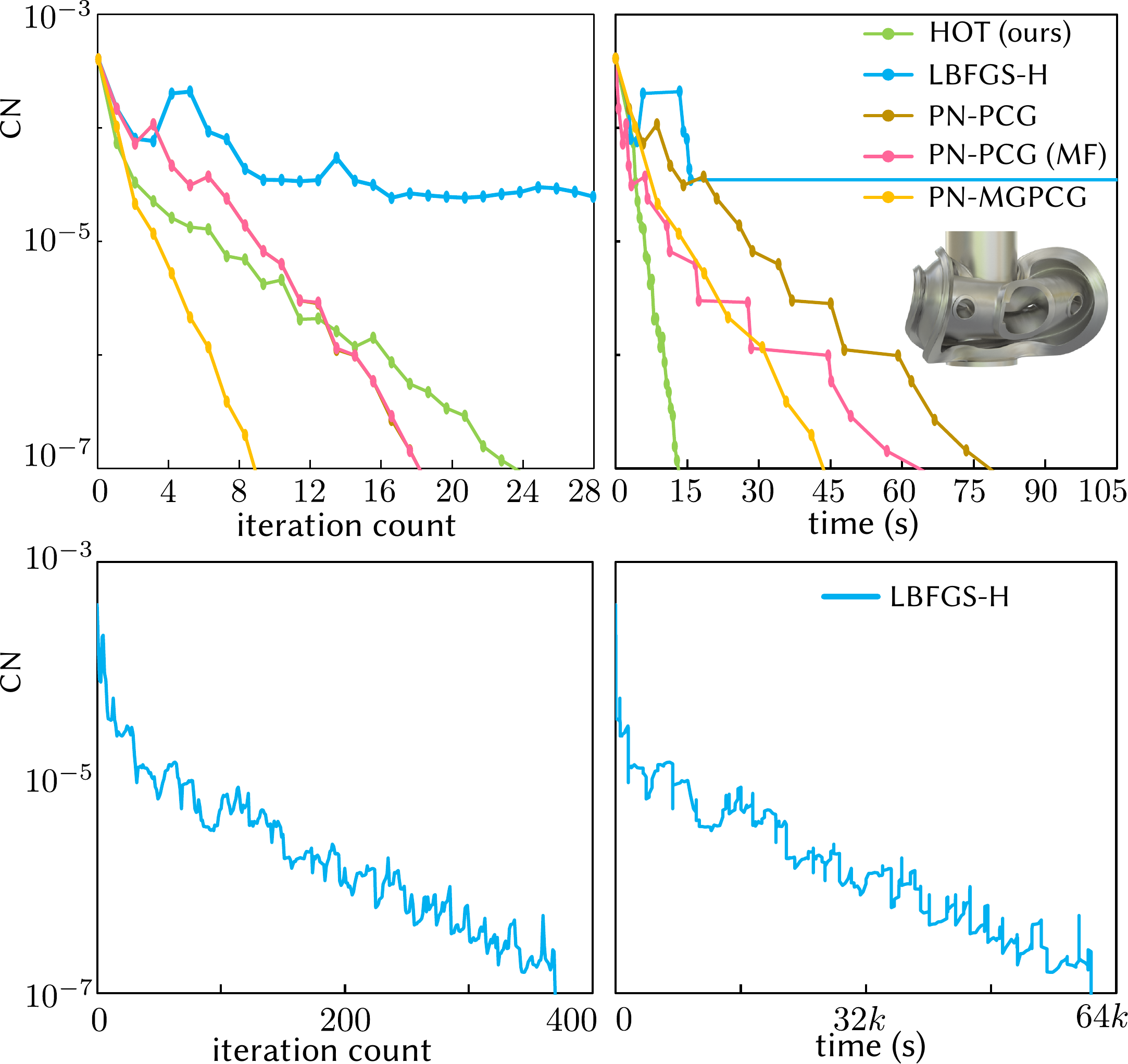}
    \caption{\textbf{Convergence comparisons.} Top left: the iteration counts for the Wheel example w.r.t. CN of different methods are visualized. Here PN-MGPCG demonstrates best convergence. Top right: total simulation times of all methods w.r.t. CN are plotted; here HOT, with low per-iteration cost obtains superior performance across all methods. Bottom: in this extreme deformation, high-stiffness example LBFGS-H converges at an extremely slow rate.}
    \label{fig:CN}
\end{figure}

\paragraph{Ablation Study}
In Fig.\ \ref{fig:CN} we compare convergence rates and timings across methods for a single time step of the Wheel example. In terms of convergence rate, we see in the top plot, that PN-MGPCG obtains the best convergence, while HOT, PN-PCG and PN-PCG(MF) converge similarly. Then, in this challenging example, LBFGS-H struggles to reach even a modest tolerance as shown in the extension in the bottom plots of Fig.\ \ref{fig:CN}.

However, for overall simulation time, HOT outperforms all three variations of PN and LBFGS-H due to its much lower per-iteration cost. 
PN-MGPCG, although with the best convergence rate, falls well behind HOT and only behaves slightly better than the two PN-PCG flavors, as the costly reconstruction of the multigrid hierarchy as well as the stiffness matrix is repeated in each Newton iteration.
LBFGS-H then struggles where we observe that many linear solves well-exceed the PCG iteration cap at $10,000$. 
At the bottom of Fig. \ref{fig:CN}, we see that LBFGS-H eventually converges after $400$ outer iterations.
Here, it appears that the diagonal preconditioner at the coarsest level in HOT significantly promotes the convergence of the whole solver; in contrast, while the same preconditioner in LBFGS-H loses its efficiency at the finest level --- the system is much much larger and conditioning of the system matrix exponentially exacerbates.

\begin{figure}[b]
    \centering
    \includegraphics[draft=\mydraft,width=\linewidth]{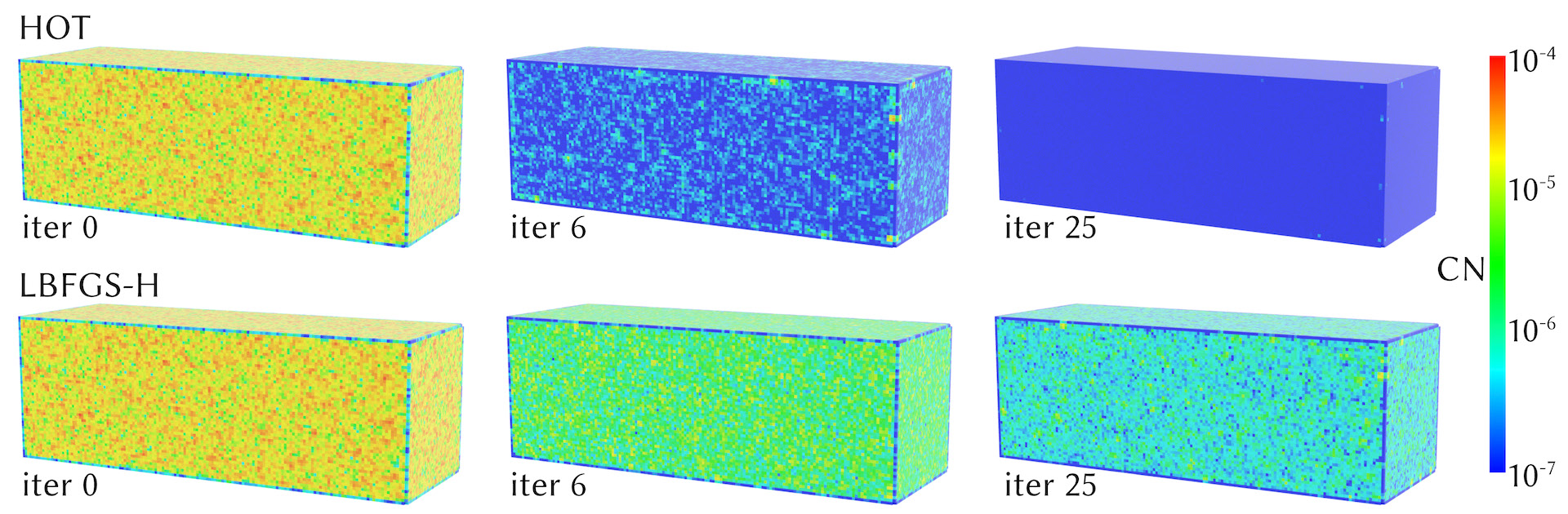}
    \caption{\textbf{Convergence visualization on stretched box.} A soft box deforms as its deformation gradient was initialized to some diagonal matrix with the diagonal entries randomly sampled in $[0.7, 1.3]$. The nodal characteristic norms of different iterations in the first time step are visualized on the rest shape. Here HOT quickly removes low-frequency errors and converges in 25 iterations, while LBFGS-H converges in 106 iterations, removing both low- and high-frequency errors simultaneously.}
    \label{fig:err}
\end{figure}

\paragraph{Visualization of Convergence}
In Fig.\ \ref{fig:err} we visualize the progressive convergence of HOT and LBFGS-H w.r.t. the CN-scaled nodal residuals for the stretched box example. 
Here HOT quickly smooths out low-frequency errors as in iteration 6 the background color of the box becomes blue (small error) and the high-frequency errors are progressively removed until HOT converges in iteration 25.
For LBFGS-H, both the low- and high-frequency errors are simultaneously removed slowly and it takes LBFGS-H 106 iterations to converge.

\begin{figure}[t]
    \centering
    \includegraphics[draft=\mydraft,width=\linewidth]{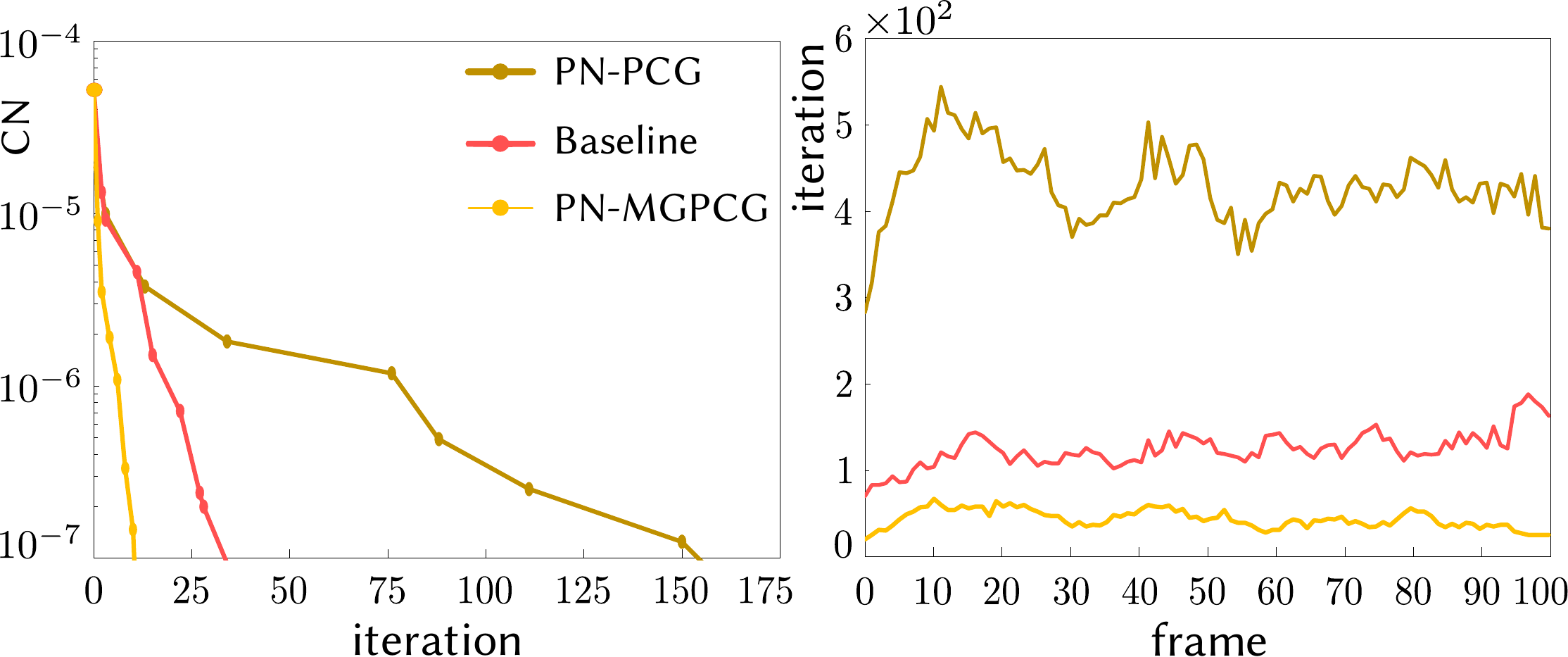}
    \caption{\textbf{Comparison to the baseline geometric multigrid.} Left: CG iteration counts in one of the time steps of the ArmaCat example w.r.t. CN of all methods. Right: per frame CG iteration counts. Here the convergence of PN-MGPCG when using the baseline geometric multigrid is worse than using our node embedding multigrid but better than Jacobi preconditioned PN-PCG. Moreover, timings of PN-PCG and PN-MGPCG highly overlaps, and they are both $3 \times$ faster than the baseline overall. }
    \label{fig:baseline}
\end{figure}

\paragraph{Comparison to the baseline geometric multigrid} \label{subsec:baseline}
As discussed above, building geometric multigrid directly from particle quadratures generally obtains essentially no speedup for coarser matrices and DoF mismatch.
We compare to this baseline geometric multigrid on the ArmaCat example (Fig.\ \ref{fig:stiffness}) by utilizing both multigrids in a PN-MGPCG time integrator.
As we see in the top plot of Fig.\ \ref{fig:baseline}, geometric multigrid effectively achieves 5$\times$ faster convergence than PN-PCG with Jacobi preconditioner, but still less effective than our 10$\times$ speedup in this specific time step.
Then the bottom plot shows that this convergence relation among all three candidates remains consistent throughout the animation of the ArmaCat example. However, in very few cases (e.g. Boards) we occasionally observed that baseline geometric multigrid preconditioned PN-MGPCG converges even more slowly than Jacobi preconditioned PN-PCG. 

Then we compare HOT to applying GMG in LBFGS (LBFGS-GMG, see Table 2 in the supplemental document). We see that the convergence of LBFGS-GMG is orders of magnitude slower than HOT for all scenes containing stiff materials.
For the two scenes with soft materials only (Sauce and Faceless), even if convergence is only slightly slower than HOT, the timing is more than $2\times$ slower, which further demonstrate the inefficiency of the multigrid operations in GMG.

\subsection{Varying Material Stiffness}

\begin{figure}[t]
    \centering
    \includegraphics[draft=\mydraft,width=\linewidth]{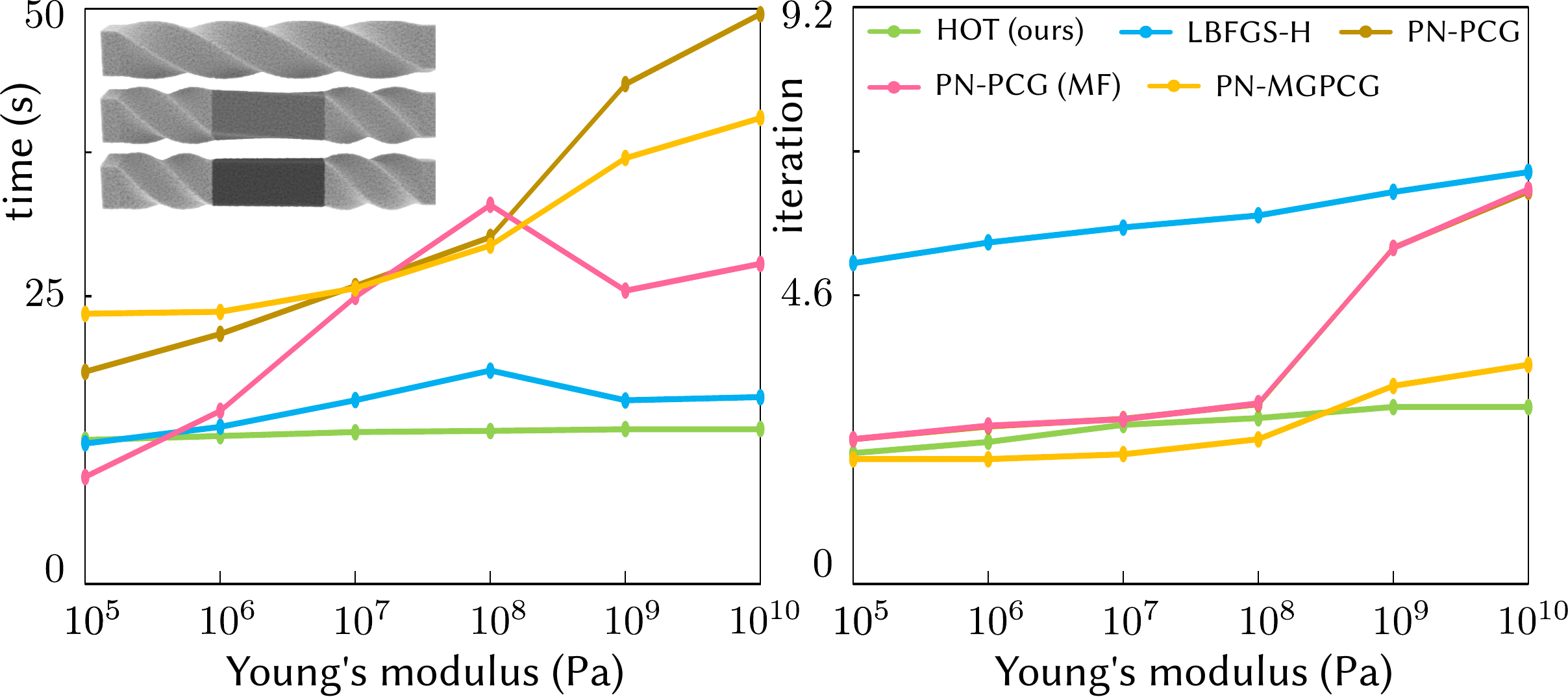}
    \caption{\textbf{Convergence and performance consistency for increasing stiffness.} Twisting a multimaterial bar we keep ends with fixed Young's moduli and progressively increase Young's for  the middle segment. Across increasing stiffness HOT exhibits the best consistency w.r.t. both  iteration count and the overall simulation time.}
    \label{fig:wedge}
\end{figure}

Finally, to again consider consistency, we compare the convergence and overall performance of all the five methods on the same simulation set-up as we progressively increase the Young's moduli of a material. Here we form a bar composed of three materials (see inset of Fig.\ref{fig:wedge}). The two bar ends are kept with a fixed constant Young's modulus of $10^5$Pa across all simulations in the experiment. We then progressively increase the Young's modulus of the middle bar from $10^5$Pa up to $10^{10}$Pa. 

In the bottom plot of Fig.\ \ref{fig:wedge}, we see that HOT maintains a low and close to flat growth in iterations for increasing stiffness with PN-MGPCG demonstrating only a modestly greater iteration growth for stiffness increase. When we consider computation time however, the modest growth in iterations for PN-MGPCG translates into much less efficient simulations as stiffness increases, due to PN-MGPCG's much greater per-iteration cost. Here, despite greater iteration growth, L-BFGS-H does better for scaling to greater stiffness due to its lower per-iteration cost. However, HOT with both a close-to-flat growth in iterations and low per-iteration cost maintains consistent behavior and generally best runtime performance with respect to increasing stiffness.  


\subsection{Ablation study on HOT's kernel}
Since our multigrid can be constructed using either linear or quadratic B-splines for node embedding which potentially leads to a trade-off between convergence and per-iteration cost because of the resulting coarser level shape functions, we here use an ablation study on the two design choices to back up our decision on using linear kernels.
As shown in Table 2 from the supplemental document, HOT with linear embedding performs equally well on convergence compared to using quadratic embedding. In a few cases e.g. ``Twist'' and ``Boxes'', linear embedding converges much faster.
This is reasonable as we can see in Fig.\ \ref{fig:our_kernel}, the resulting shape functions on the coarser level when using linear or quadratic node embedding do not have significant differences.
But because linear embedding leads to much sparser coarse systems, it is much faster on timing than quadratic embedding.

\section{Conclusions and Future Work} \label{sec:discussion}
Starting from node embedding we derive, to our knowledge, the first MPM Galerkin multigrid. The resulting method is consistent with particle quadrature geometric multigrid while still providing efficiency in construction and automatically handling boundary conditions.
We then build HOT to utilize our multigrid as an inner initializer within L-BFGS to avoid the repeated expensive matrix reconstruction costs required in traditional PN-MGPCG. Together with efficient curvature updates we ensure fast yet inexpensive convergence.
HOT accelerates implicit MPM time-stepping to a wide range of important material behaviors, achieving up to $10\times$ speedup compared to the state-of-the-art methods and shines among an exhaustive set of variations. We hope that this will enable further research to leverage HOT-type hierarchies to address both spatial and temporal discretization limitations in current MPM pipelines. 

Semi-implicit plasticity is a limitation of HOT. For current HOT simulations with plasticity, we have not encountered instabilities. Nevertheless, adding plasticity return mapping as a constraint within HOT's stable implicit time-step optimization is an exciting and challenging future direction to explore.
Another challenging but interesting extension would be to incorporate additional physical constraints in our minimization framework. One particularly useful case is volume preservation for e.g. simulating stiff granular materials. 
Likewise, although our proposed inexact criteria resolves heterogeneous materials, we believe that it can and should be further improved. 
Currently HOT offers consistent and performant MPM simulations to gain the advantages of implicit timestepping without need for parameter tuning. We look forward to further extensions, such as those above, and its application as a tool kit to accelerate the applications of MPM for the physical sciences, animation, engineering and beyond.

On the implementation side, the construction of the finest level system matrix is one of the bottlenecks for HOT.
In our code it is realized with a scattering scheme, which suffers from cache misses.
Therefore, exploring the performance potential of alternative gathering schemes for building the stiffness matrices can be a meaningful future work.


\section*{Acknowledgement}
We would like to thank Hannah Bollar for narrating the video, Joshuah Wolper for proofreading, and the anonymous reviewers for their valuable comments. This work was supported in part by NSFC (61972341, 61972342, 61732015, 61572423), NSF Grants IIS-1755544 and CCF-1813624, DOE ORNL subcontract 4000171342, a gift from Adobe Inc., NVIDIA GPU grants, and Houdini licenses from SideFX.

\bibliographystyle{ACM-Reference-Format}
\bibliography{references}



\clearpage


\section*{Supplementary material}

# Hierarchical Optimization Time Integration for CFL-rate MPM Stepping

## 1 Benchmark Summary Table

For performance and convergence comparison, we put timing and iteration results in the following two tables. *avg time* measures average absolute cost (seconds) per playback frame, *total* measures the HOT speedup factor of the wall clock time for the entire rendered animation sequence, *max* records the maximum speedup factor HOT achieved on a simulated (and rendered) at 24Hz frame, *avg iter* (or *iter*) measures the average number of Newton or quasi-Newton outer iterations (per method) required per frame to achieve the requested accuracy. Each example is run for all methods on the same machine. Machines employed per example: *Twist*, *Chain* and *Wheel*: Intel Core i7-7700K; all other examples are run on an Intel Core i7-8700K. Both machines has 64GB memory. Cat Young's modulus values are [†]$10^6$ and [‡]$10^9$ respectively. [*] indicates that the examples could not finish in reasonable time, and was manually terminated.

Table 1: **Newton's Method Timings:** Here we summarize statistics across all benchmark examples using Newton's methods (including the previous state-of-the-art Gast15 [1] in comparison with HOT. Here, Gast15 method consistently adopts 1e-3 as the outer tolerance for all examples, which is the maximum that guarantees artifact-free results.

| Example | HOT | | Gast15(MF) | | | | PN-PCG | | | PN-PCG(MF) | | | PN-MGPCG | | |
|---|---|---|---|---|---|---|---|---|---|---|---|---|---|---|---|
| | avg time | avg iter | avg time | total | | iter | total | max | iter | total | max | iter | total | max | iter |
| Twist | 77.73 | 13.49 | *2308.70 | *29.70× | *19.33 | | 4.65× | 8.17× | 11.14 | 4.73× | 9.57× | 11.14 | 6.79× | 9.85× | 5.42 |
| Boxes | 129.81 | 5.76 | *10142.33 | *78.13× | *12.14 | | 3.59× | 9.29× | 7.21 | 3.73× | 9.19× | 7.21 | 3.57× | 7.91× | 3.94 |
| Donut | 121.19 | 27.76 | *1150.41 | *9.49× | *15.68 | | 1.98× | 7.61× | 9.07 | 1.98× | 9.39× | 9.07 | 10.67× | 17.97× | 4.68 |
| [†]ArmaCat | 32.55 | 6.22 | 62.78 | 1.93× | 8.60 | | 3.41× | 4.53× | 7.03 | 1.22× | 1.79× | 7.03 | 3.21× | 3.87× | 4.69 |
| [‡]ArmaCat | 36.61 | 8.72 | 324.77 | 8.87× | 13.94 | | 4.19× | 6.28× | 8.40 | 2.02× | 3.78× | 8.40 | 3.42× | 3.43× | 5.38 |
| Chain | 98.78 | 5.55 | *766.47 | *7.76× | *9.84 | | 5.79× | 11.99× | 6.04 | 1.98× | 6.85× | 6.04 | 4.02× | 8.69× | 3.42 |
| Boards | 105.99 | 3.72 | 296.43 | 2.80× | 2.74 | | 2.95× | 5.77× | 3.11 | 1.73× | 7.39× | 3.11 | 2.51× | 4.76× | 2.402 |
| Wheel | 44.38 | 8.56 | *39447.37 | *888.85× | *54.5 | | 4.64× | 5.93× | 8.42 | 5.76× | 6.74× | 8.42 | 3.58× | 4.88× | 5.96 |
| Faceless | 3.49 | 6.44 | 2.84 | 0.81× | 2.09 | | 2.06× | 5.74× | 4.49 | 1.68× | 7.05× | 4.49 | 2.25× | 6.42× | 3.81 |
| Sauce | 13.11 | 4.54 | 10.42 | 0.79× | 3.21 | | 2.22× | 5.77× | 4.93 | 1.05× | 2.69× | 4.93 | 2.26× | 2.82× | 3.18 |

Table 2: **HOT Timing Comparisons:** Here we summarize statistics across all benchmark examples and methods that partly resemble our HOT. Compared to HOT, both LBFGS-GMG and LBFGS-H use LBFGS as the quasi-Newton solver but with different initializers, i.e. baseline particle quadrature multigrid for LBFGS-GMG and inexact PCG for LBFGS-H. PN-MGPCG adopts the same multigrid formulation from HOT yet a different nonlinear optimization method. HOT-quadratic is the derivation of HOT whose multigrid is built according to quadratic kernel rather than linear kernel. As a result, all these alternatives are much less efficient than HOT in general.

| Example | HOT | | HOT-quadratic | | | LBFGS-GMG | | LBFGS-H | | | PN-MGPCG | | |
|---|---|---|---|---|---|---|---|---|---|---|---|---|---|
| | avg time | avg iter | total | max | iter | total | iter | total | max | iter | total | max | iter |
| Twist | 77.73 | 13.49 | 7.10× | 86.42× | 51.24 | *186.93× | *1234.94 | 4.12× | 9.53× | 20.45 | 6.79× | 9.85× | 5.42 |
| Boxes | 129.81 | 5.76 | 2.54× | 4.60× | 9.61 | *61.41× | *296.56 | 2.39× | 8.84× | 6.78 | 3.57× | 7.91× | 3.94 |
| Donut | 121.19 | 27.76 | 2.18× | 4.59× | 32.81 | *85.38× | *1182.52 | 4.79× | 2.63× | 16.42 | 10.67× | 17.97× | 4.68 |
| [†]ArmaCat | 32.55 | 6.22 | 2.01× | 2.09× | 6.17 | 2.93× | 18.70 | 0.94× | 1.72× | 8.09 | 3.21× | 3.87× | 4.69 |
| [‡]ArmaCat | 36.61 | 8.72 | 1.94× | 3.18× | 8.67 | *201.56× | *709.05 | 1.37× | 2.45× | 8.95 | 3.42× | 3.43× | 5.38 |
| Chain | 98.78 | 5.55 | 2.91× | 5.77× | 4.54 | *7.59× | *166.57 | 1.92× | 5.83× | 6.26 | 4.02× | 8.69× | 3.42 |
| Boards | 105.99 | 3.72 | 2.83× | 4.09× | 3.56 | 4.98× | 39.87 | 2.01× | 5.13× | 6.252 | 2.51× | 4.76× | 2.402 |
| Wheel | 44.38 | 8.56 | 2.27× | 2.49× | 7.77 | *2403.47× | *5817 | *51.62× | *217.75× | *16.36 | 3.58× | 4.88× | 5.96 |
| Faceless | 3.49 | 6.44 | 1.80× | 2.20× | 6.56 | 6.12× | 9.64 | 1.03× | 1.31× | 9.19 | 2.25× | 6.42× | 3.81 |
| Sauce | 13.11 | 4.54 | 1.97× | 2.82× | 4.56 | 2.86× | 6.13 | 0.92× | 5.45× | 7.76 | 2.26× | 2.82× | 3.18 |

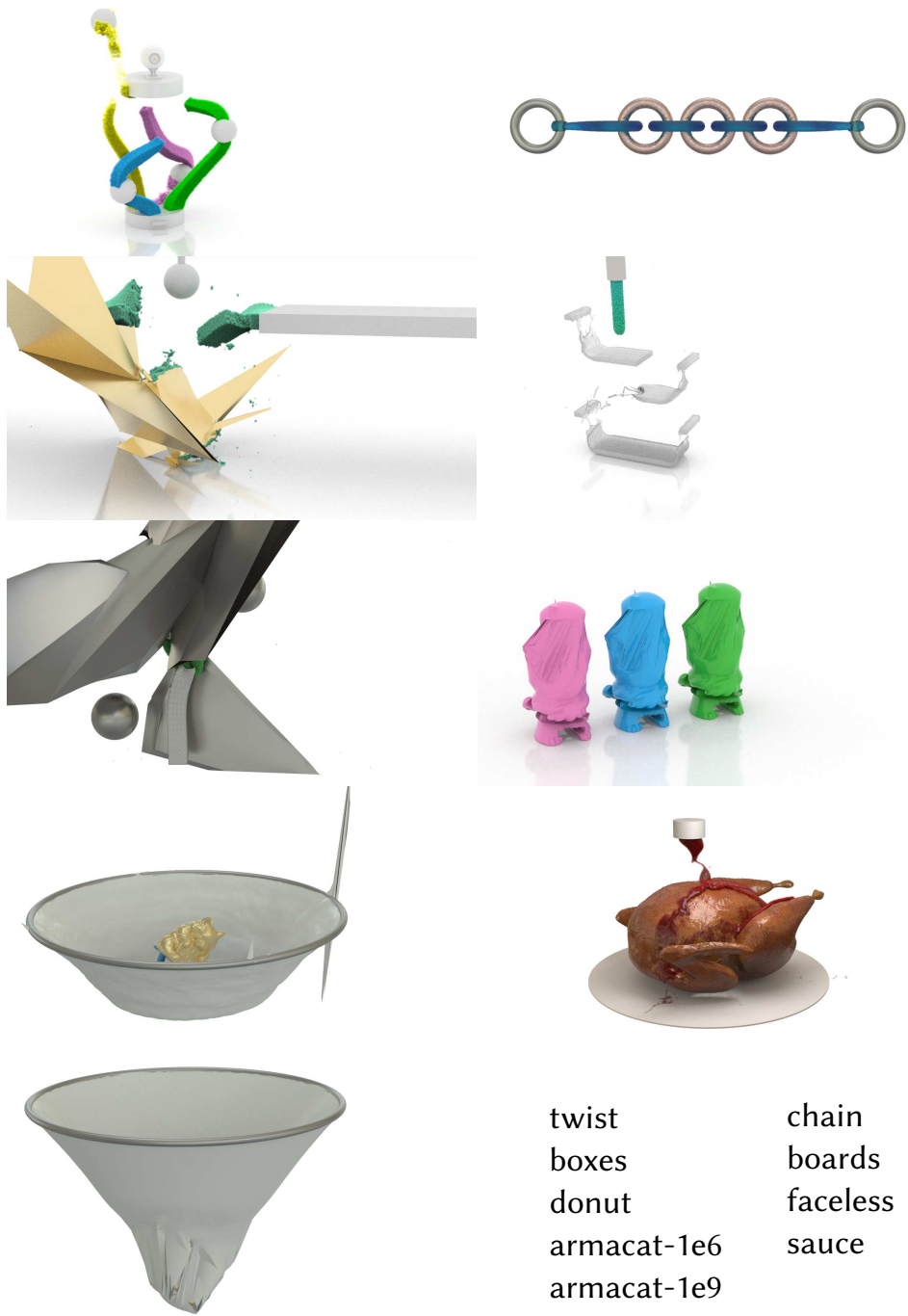

Figure 1: **Artifacts**. Various scales of explosions can be observed among *twist*, *boxes*, *donut*, and [†]*armacat(1e6)*. Artificial softening occurs in [‡]*armacat(1e9)*, *boards*, *faceless* and *sauce*. In *chain*, rings in the middle are not pulled from each other under forces from both two sides.

## 2  Gast15 Failed Cases

In this section, we demonstrate all failed results (Figure 1) generated from the previous state-of-the-art Gast15 [1] using the same tolerance $10^2$. These models exhibit obvious artifacts of all kinds due to the inappropriate tolerance setting in each example except for wheel. The largest tolerance that produce artifact-free results varies across examples and this inconsistency brings significant inconvenience to the setup of a new simulation, even worse for cases where material properties change throughout the simulation.



\end{document}